\documentclass[reprint,pre]{revtex4-1}

\usepackage{amsmath}
\usepackage{graphicx}

\usepackage [novolumeabbr] {unitsdef} 

\usepackage{cancel}
\usepackage{deleq}

\DeclareMathOperator{\Tr}{Tr}   
\DeclareMathOperator{\cc}{c.c.} 
\DeclareMathOperator{\Real}{Re}   
\DeclareMathOperator{\Imag}{Im}   
\DeclareMathOperator{\D}{D} 

\newunit{\torr}{Torr}	
\newunit{\nanovolt}{nV} 
\newunit{\microvolt}{\mu V} 
\newunit{\Kelvin}{\degree\kelvin} 

\begin{document}
\title {Nonlinear damping in a micromechanical oscillator}

\author{Stav~Zaitsev}
\email{zzz@tx.technion.ac.il}

\author{Oleg~Shtempluck}

\author{Eyal~Buks}
\affiliation{Electrical Engineering Department, Technion - Israel Institute of Technology, Haifa, Israel 32000}

\author{Oded~Gottlieb}
\affiliation{Mechanical Engineering Department, Technion - Israel Institute of Technology, Haifa, Israel 32000}

\begin{abstract}

Nonlinear elastic effects play an important role in the dynamics of microelectromechanical systems (MEMS). A Duffing oscillator is widely used as an archetypical model of mechanical resonators with nonlinear elastic behavior. In contrast, nonlinear dissipation effects in micromechanical oscillators are often overlooked. In this work, we consider a doubly clamped micromechanical beam oscillator, which exhibits nonlinearity in both elastic and dissipative properties. The dynamics of the oscillator is measured in both frequency and time domains and compared to theoretical predictions based on a Duffing-like model with nonlinear dissipation. We especially focus on the behavior of the system near bifurcation points. The results show that nonlinear dissipation can have a significant impact on the dynamics of micromechanical systems. To account for the results, we have developed a continuous model of a geometrically nonlinear beam-string with a linear Voigt-Kelvin viscoelastic constitutive law, which shows a relation between linear and nonlinear damping. However, the experimental results suggest that this model alone cannot fully account for all the experimentally observed nonlinear dissipation, and that additional nonlinear dissipative processes exist in our devices.
\end{abstract}

\maketitle

\section{Introduction}
\label{sec:introduction}

The field of micro-machining is forcing a profound redefinition of the nature and attributes of electronic devices. This technology allows fabrication of a variety of on-chip fully integrated micromechanical sensors and actuators with a rapidly growing range of applications. In many cases, it is highly desirable to shrink the size of mechanical elements down to the nano-scale \cite{Turner_et_al_98, Roukes_NEMS_01, Roukes_NEMS_00, Husain_et_al_03}. This allows enhancing the speed of operation by increasing the frequencies of mechanical resonances and improving their sensitivity as sensors. Furthermore, as devices become smaller, their power consumption decreases and the cost of mass fabrication can be significantly lowered. Some key applications of microelectromechanical systems (MEMS) technology include magnetic resonance force microscopy (MRFM) \cite{Sidles_et_al_95, Rugar_et_al_04} and mass-sensing \cite{Zhang_et_al_02, Ekinci_et_al_04, Ekinci_et_al_04a, Ilic_et_al_04}. Further miniaturization is also motivated by the quest for mesoscopic quantum effects in mechanical systems \cite{Blencowe_04, Knobel&Cleland_03, Lahaye_et_al_04, Schwab_et_al_00, Buks&Roukes_01a, Buks&Roukes_01b, Schwab&Roukes_05, Aspelmeyer&Schwab_08}.

Nonlinear effects are of great importance for micromechanical devices. The relatively small applied forces needed for driving a micromechanical oscillator into the nonlinear regime are usually easily accessible \cite{Kozinsky_et_al_07}. Thus, a variety of useful applications such as frequency synchronization \cite{Cross_et_al_04}, frequency mixing and conversion \cite{Erbe_et_al_00, Reichenbach_et_al_05}, parametric and intermodulation amplification \cite{Almog_et_al_06a}, mechanical noise squeezing \cite{Almog_et_al_06b}, stochastic resonance \cite{Almog_et_al_07}, and enhanced sensitivity mass detection \cite{Buks&Yurke_06} can be implemented by applying modest driving forces. Furthermore, monitoring the displacement of a micromechanical resonator oscillating in the linear regime may be difficult when a displacement detector with high sensitivity is not available. Thus, in many cases the nonlinear regime is the only useful regime of operation.

Another key property of systems based on mechanical oscillators is the rate of damping. For example, in many cases the sensitivity of NEMS sensors is limited by thermal fluctuation \cite{Cleland&Roukes_02, Ekinci_et_al_04}, which is related to damping via the fluctuation dissipation theorem. In general, micromechanical systems suffer from low quality factors $Q$ relative to their macroscopic counterparts \cite{Roukes_NEMS_00, Yasumura_et_al_00, Ono_et_al_03}. However, very little is currently known about the underlying physical mechanisms contributing to damping in these devices. A variety of different physical mechanisms can contribute to damping, including bulk and surface defects \cite{Liu_et_al_99, Harrington_et_al_00}, thermoelastic damping \cite{Lifshitz&Roukes_00, Houston_et_al_02}, nonlinear coupling to other modes, phonon-electron coupling, clamping loss \cite{Lifshitz_02, Wilson-Rae_08}, interaction with two level systems \cite{Remus_et_al_09}, etc. Identifying experimentally the contributing mechanisms in a given system can be highly challenging, as the dependence on a variety of parameters has to be examined systematically \cite{Jaksic&Boltezar_02, Popovic_et_al_95, Zhang_et_al_02a, Zhang_et_al_03}.

The archetypical model used to describe nonlinear micro- and nanomechanical oscillators is the Duffing oscillator \cite{Nayfeh_Mook_book_95}. This model has been studied in great depth \cite{Dykman&Krivoglaz_84, Landau&Lifshitz_Mechanics, Nayfeh_book_81, Nayfeh_Mook_book_95}, and special emphasis has been given to the dynamics of the system near the bifurcation points \cite{Arnold_book_88, Strogatz_book_94, Chan_et_al_08, Dykman_et_al_04, Yurke&Buks_06, Buks&Yurke_06a}.

In order to describe dissipation processes, a linear damping model is usually employed, either as a phenomenological ansatz, or in the form of linear coupling to thermal bath, which represents the environment. However, nonlinear damping is known to be significant at least in some cases. For example, the effect of nonlinear damping for the case of strictly dissipative force, being proportional to the velocity to the n'th power, on the response and bifurcations of driven Duffing \cite{Ravindra&Mallik_94a, Ravindra&Mallik_94b, Trueba_et_al_00, Baltanas_et_al_01} and other types of nonlinear oscillators \cite{Nayfeh_Mook_book_95, Sanjuan_99, Trueba_et_al_00} has been studied extensively. Also, nonlinear damping plays an important role in parametrically excited mechanical resonators \cite{Lifshitz&Cross_08} where without it, solutions will grow without bound \cite{Nayfeh_Mook_book_95, Gutschmidt&Gottlieb_09}.

In spite of the fact that a massive body of literature exists which discusses the nonlinear elastic effects in micro- and nanomechanical oscillators as well as the consequences of nonlinear damping, the quantitative experimental data on systems with nonlinear damping, especially those nearing bifurcation points, remains scarce. Furthermore, such systems impose special requirements on the experiment parameters and procedures, mainly due to the very slow response times near the bifurcation points. Straightforward evaluation of these requirements by simple measurements can facilitate accurate data acquisition and interpretation.

In the present paper we study damping in a micromechanical oscillator operating in the nonlinear regime excited by an external periodic force at frequencies close to the mechanical fundamental mode. We consider a Duffing oscillator nonlinearly coupled to a thermal bath. This coupling results in a nonlinear damping force proportional to the velocity multiplied by the displacement squared. As will be shown below, this approach is equivalent to the case where the damping nonlinearity is proportional to the velocity cubed \cite{Lifshitz&Cross_03}. In conjunction with a linear dissipation term, it has also been shown to describe an effective quadratic drag term \cite{Bikdash_et_al_94}.

We find that nonlinear damping in our micromechanical oscillators is non-negligible, and has a significant impact on the oscillators' response. Furthermore, we develop a theoretical one-dimensional model of the oscillator's behavior near the bifurcation point \cite{Arnold_book_88, Dykman&Krivoglaz_84}. Most of the parameters that govern this behavior can be estimated straightforwardly from frequency response measurements alone, not requiring exact measurement of oscillation amplitudes.  Measuring these parameters under varying conditions provides important insights into the underlying physical mechanisms \cite{Gottlieb&Feldman_97, Dick_et_al_06}.

We use our results to estimate different dynamic parameters of an experimentally measured micromechanical beam response, and show how these estimations can be used to increase the accuracy of experimental measurements and to estimate measurement errors. The main source of error is found to be the slowing down behavior near the bifurcation point, also known as the saddle node "ghost" \cite{Strogatz_book_94}. We also investigate the possibility of thermal escape of the system from a stable node close to the bifurcation point \cite{Dykman&Krivoglaz_84, Dykman_et_al_04, Aldridge&Cleland_05, Chan_et_al_08} and find that the probability of this event in our experiments is negligible.

Finally, we propose and analyze a continuum mechanics model of our micromechanical oscillator as a planar, weakly nonlinear strongly pretensioned, viscoelastic beam-string. The analysis of this model illustrates a possible cause for non negligible nonlinear damping as observed in the experiment.

\section{Experimental setup}
\label{sec:exp_setup}

\begin{figure} [htb]
        \includegraphics [width=3.4in] {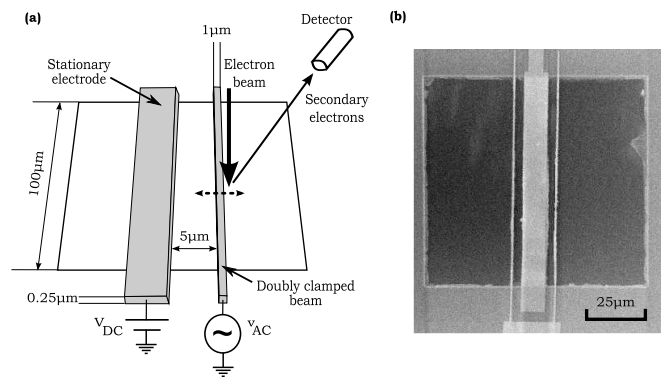}
        \caption {A typical device consists of a suspended doubly clamped narrow beam (length 200\micm, width 1-0.25\micm, and thickness 0.2\micm) and a wide electrode. The excitation force is applied as voltage between the beam and the electrode. \textbf{(a)} Experimental setup and typical sample's dimensions. The direction of the vibration of the micromechanical beam is denoted by dotted arrow. \textbf{(b)} SEM micrograph of a device with one wide electrode and two narrow doubly clamped beams.}
        \label{fig:exp_setup}
\end{figure}

For the experiments we employ micromechanical oscillators in the form of doubly clamped beams made of PdAu (see Fig.~\ref{fig:exp_setup}). The device is fabricated on a rectangular silicon-nitride membrane (side length 100-200\micm) by the means of electron beam lithography followed by thermal metal evaporation. The membrane is then removed by electron cyclotron resonance (ECR) plasma etching, leaving the doubly clamped beam freely suspended. The bulk micro-machining process used for sample fabrication is similar to the one described in \cite{Buks&Roukes_01b}. The dimensions of the beams are: length 100-200\micm, width 0.25-1\micm and thickness 0.2\micm, and the gap separating the beam and the electrode is 5-8\micm. 

Measurements of all mechanical properties are done \textit{in-situ} by a scanning electron microscope (SEM) (working pressure $10^{-5}\torr$), where the imaging system of the microscope is employed for displacement detection \cite{Buks&Roukes_01b}. Some of the samples were also measured using an optical displacement detection system described elsewhere \cite{Almog_et_al_06b}. Driving force is applied to the beam by applying a voltage to the nearby electrode. With a relatively modest driving force, the system is driven into the region of nonlinear oscillations \cite{Buks&Roukes_01b, Buks&Roukes_02}.

\begin{figure} [htb]
    
        \includegraphics [width=3.4in] {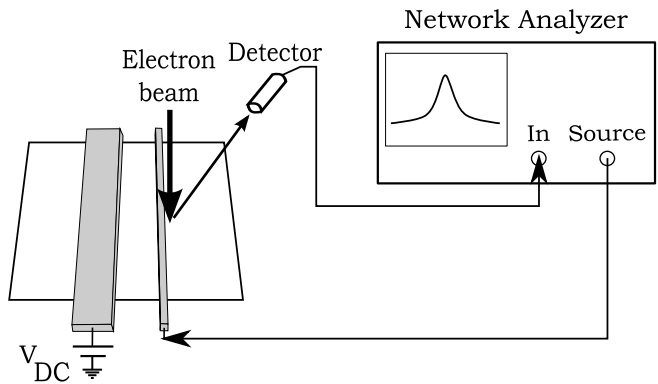}
        \caption {Network analyzer is used for frequency domain measurements. If the system is excited into a bistable regime, special care should be taken to ensure accurate measurement near bifurcation points, as discussed in Sec.~\ref{sec:sweep_cond}.}
        \label{fig:exp_setup_NA}
    
\end{figure}

\begin{figure} [htb]
    
        \includegraphics [width=3.4in] {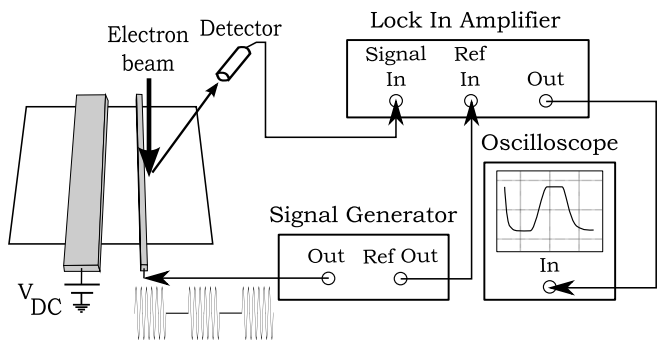}
        \caption {Lock-in amplifier is employed for time domain measurements. The oscillator is excited at a single frequency. The amplitude of the excitation is modulated by a square wave, effectively turning the excitation on and off 20-50 times per second. Such bursting excitation is used to measure the ringing down of the slow envelope in the time domain.}
        \label{fig:exp_setup_Lockin}
    
\end{figure}

We use a network analyzer for frequency domain measurements, as shown in Fig.~\ref{fig:exp_setup_NA}. For time domain measurements of the slow varying envelope we employ a lock-in amplifier, connected as show in Fig.~\ref{fig:exp_setup_Lockin}. The mechanical oscillator is excited by a monochromatic wave, whose amplitude is modulated by a square wave with low frequency (20-50\hertz). This results in bursting excitation, which allows measurement of ring-down behavior in time domain. The lock-in amplifier is locked to the excitation frequency, and measures the amplitude of the slow envelope of the oscillator's response. The lock-in amplifier time constant should be much smaller than the ring down time, which is governed by dissipation in the micromechanical system. Typically, in our experiments, the time constant is 100\microsecond and the characteristic ring down time is 10\millisecond.

The displacement detection scheme described above is not exactly linear, because the amount of the detected secondary electrons or reflected light is not strictly proportional to the mechanical oscillator amplitude, but merely a monotonic function of the latter. Nonuniform distribution of primary electrons or light power in the spot increases this nonlinearity even further. Thus, some distortion in the measured response amplitude is introduced.

\section{Theory}

	\subsection{Equation of motion}

We excite the system close to its fundamental mode. Ignoring all higher modes allows us to describe the dynamics using a single degree of freedom $x$.

In the main part of this study, no assumptions are made about the source of linear and nonlinear dissipation. The energy dissipation is modeled phenomenologically by coupling the micromechanical oscillator to a thermal bath consisting of harmonic oscillators \cite{Ullersma_66a, Ullersma_66b, Caldeira&Leggett_83, Hanggi_97}. Physically, several processes may be responsible for mechanical damping \cite{Cleland&Roukes_02, Lifshitz_02, Mohanty_et_al_02, Yasumura_et_al_00, Zener_book_48}, including thermoelastic effects \cite{Stievater_et_al_07,  Houston_et_al_02, Lifshitz&Roukes_00}, friction at grain boundaries \cite{Ke_47b}, bulk and surface impurities \cite{Ono&Esashi_05, Zolfagharkhani_et_al_05, Ono_et_al_03}, electrical losses, clamping loss \cite{Geller&Varley_05, Cross&Lifshitz_01, Wilson-Rae_08}, etc. We also regard the linear and nonlinear damping constants as independent of one another, although they probably result from same physical processes. In Sec.~\ref{sec:nonlin_damping_source} we consider one possible model connecting the linear and nonlinear dissipation coefficients, and compare its predictions to experimental data.

The Hamiltonian of the system, which includes the mechanical beam and thermal bath modes coupled to it, is
\begin{equation}
    \mathcal{H}=\mathcal{H}_m+\mathcal{H}_b+\mathcal{H}_i,
	\label{eq:Hamiltonian}%
\end{equation}
where
\begin{align*}
    \mathcal{H}_m&=\frac{p^2}{2m}+\tilde U(x)+\mathcal{E}_\text{cap}(x,t),\\
	\mathcal{H}_b&=\sum_{b}\left(\frac{p_b^2}{2m_b}+\frac{1}{2}m_b\omega_b^2 q_b^2\right),\\
	\mathcal{H}_i&=\sum_{b}\Gamma(x,\omega_b)q_b,
\end{align*}
describe the micromechanical beam, the thermal bath, and the interaction between them, respectively. Here, $m$ is the effective mass of the fundamental mode of the micromechanical beam, and $p$ and $x$ are the effective momentum and displacement of the beam. Also, $\tilde U(x)$ is  the elastic potential, and $\mathcal{E}_\text{cap}(x,t)=C(x)V(t)^2/2$ is the capacitive energy, where $C(x)=C_{0}/(1-x/d)$ is the displacement dependent capacitance, $d$ is the gap between the electrode and the beam, and $V(t)$ is the time dependent voltage applied between the electrode and the micromechanical beam. The sum $\sum_{b}$ denotes summing over all relevant thermal bath modes, while $\omega_b$ is the frequency of one of the modes in the thermal bath with effective momentum $p_b$ and displacement $q_b$, and $m_b$ is the effective mass of the same mode. Finally, $\Gamma(x,\omega_b)$ is a function describing the interaction strength of each thermal bath mode with the fundamental mode of the micromechanical beam.

The equations of motion resulting from \eqref{eq:Hamiltonian} are
\begin{subequations}
	\begin{align}
		m\ddot x &= -\frac{\partial}{\partial x}\left(\tilde U(x)+\mathcal{E}_\text{cap}(x,t)\right)-\sum_{b}q_b\frac{\partial \Gamma(x,\omega_b)}{\partial x}, \label{eq:x_Hamiltonian_eq_of_motion} \\
		m_b\ddot q_b &= -m_b\omega_b^2 q_b-\Gamma(x,\omega_b). \label{eq:qb_eq_of_motion}
	\end{align}
\end{subequations}
The formal solution of \eqref{eq:qb_eq_of_motion} can be written as
\begin{multline*}
	q_b(t)=q_{b0}\cos\omega_b t+\frac{\dot q_{b0}}{\omega_b}\sin\omega_b t\\
	+\int_0^t \frac{\Gamma(x,\omega_b;\tau)}{m_b\omega_b} \sin\omega_b(\tau-t) d\tau,
\end{multline*}
or, integrating by parts,
\begin{multline}
	q_b(t)=q_{b0}\cos\omega_b t+\frac{\dot q_{b0}}{\omega_b}\sin\omega_b t \\
	+\frac{\Gamma(x,\omega_b;0)}{m_b\omega_b^2} \cos\omega_b t-\frac{\Gamma(x,\omega_b;t)}{m_b\omega_b^2}\\
	+\int_0^t \frac{\dot x(\tau)}{m_b\omega_b^2}\frac{\partial \Gamma(x,\omega_b;\tau)}{\partial x} \cos\omega_b(\tau-t) d\tau,
	\label{eq:qb_solution}
\end{multline}
where $q_{b0}$ and $\dot q_{b0}$ are the initial conditions of the thermal mode displacement and velocity, respectively; and $\Gamma(x,\omega_b;s)$ denotes the coupling strength function $\Gamma(x,\omega_b)$ evaluated at time $s$.

Substituting \eqref{eq:qb_solution} into \eqref{eq:x_Hamiltonian_eq_of_motion}, one gets
\begin{multline}
		m\ddot x+\int_0^t \mathcal{K}(x,t,\tau)\dot x(\tau) d\tau+\frac{\partial U(x)}{\partial x}\\
	=-\frac{\partial\mathcal{E}_\text{cap}(x,t)}{\partial x}+mn(t),
		\label{eq:x_integral_eq_of_motion}
\end{multline}
where $n(t)$ is the noise,
\begin{equation}
	U(x)=\tilde U(x)-\sum_b \frac{\Gamma^2(x,\omega_b)}{2m_b\omega_b^2}
	\label{eq:V_renormalized}
\end{equation}
is the renormalized potential, and
\begin{equation*}
	\mathcal{K}(x,t,\tau)=\sum_b \frac{\partial \Gamma(x,\omega_b;t)}{\partial x}\frac{\partial \Gamma(x,\omega_b;\tau)}{\partial x} \frac{\cos\omega_b(\tau-t)}{m_b\omega_b^2} 
\end{equation*}
is the memory kernel \cite{Hanggi_97, Hanggi&Ingold_04}. Also, the initial slip term given by $$\sum_b \Gamma(x,\omega_b;0) \cos\omega_b t \frac{\partial}{\partial x}\Gamma(x,\omega_b;t) /(m_b\omega_b^2),$$ has been dropped \cite{Hanggi_97}. Finally, the noise autocorrelation for an initial thermal ensemble is
\begin{equation*}
	\left<n(t)n(s) \right>=\frac{k_BT}{m^2}\mathcal{K}(x,t,s),
\end{equation*}
where $T$ is the effective temperature of the bath, and $k_B$ is the Boltzmann's constant. The last result is a particular form of the fluctuation-dissipation theorem \cite{Landau&Lifshitz_StatMechPt1, Kubo_66, Chandrasekhar_43, Klimontovich_book_95}.

We employ a nonlinear, quartic potential $U(x)=\frac{1}{2}k_1x^2+\frac{1}{4}k_3x^4$ in order to describe the elastic properties of the micromechanical beam oscillator. Assuming $\Gamma(x,\omega_b)$ to be polynomial in $x$, it can be deduced from \eqref{eq:V_renormalized} that only linear and quadratic terms in $\Gamma(x,\omega_b)$ should be taken into account \cite{Caldeira&Leggett_83, Habib&Kandrup_92}, i.e.,
\begin{equation}
	\Gamma(x,\omega_b)=g_1(\omega_b)x+\frac{1}{2}g_2(\omega_b)x^2.
	\label{eq:Gamma}
\end{equation}
The memory kernel in this case is
\begin{multline*}
	\mathcal{K}(x,t,\tau)=\sum_b \Bigl(g_1^2+g_1g_2(x(t)+x(\tau))\\
	+g_2^2x(t)x(\tau)\Bigl)\frac{\cos\omega_b(\tau-t)}{m_b\omega_b^2}.
\end{multline*}
Making the usual Markovian (short-time noise autocorrelation) approximation \cite{Ullersma_66a, Caldeira&Leggett_83, Dykman&Krivoglaz_84}, i.e., $\mathcal{K}(x,t,s)\propto\delta(t-s)$, one obtains
\begin{equation*}
	\mathcal{K}(x,t,\tau)=\left(2b_{11}+b_2 x+b_{31}x^2\right)\delta(t-\tau),
\end{equation*}
and the equation of motion \eqref{eq:x_integral_eq_of_motion} becomes
\begin{multline}
		m\ddot x+(2b_{11}+b_{31}x^2+b_{32}\dot x^2)\dot x+k_{1}x+k_{3}x^{3}\\
	=-\frac{\partial\mathcal{E}_\text{cap}(x,t)}{\partial x}+mn(t),
		\label{eq:Newton_eq}
\end{multline}
where $b_{11}$ is the linear damping constant, $b_{31}$ and $b_{32}$ are the nonlinear damping constants, $k_{1}$ is the linear spring constant and $k_{3}$ is the nonlinear spring constant.

Some clarifications regarding \eqref{eq:Newton_eq} are in order. The quadratic dissipation term $b_2x\dot x$ has been dropped  from the equation because it has no impact on the first order multiple scales analysis, which will be applied below. An additional dissipation term proportional to the cubed velocity, $b_{32}\dot x^3$, has been added artificially. Such term, although not easily derived using the analysis sketched above, may be required to describe some macroscopic friction mechanisms \cite{Nayfeh_Mook_book_95, Ravindra&Mallik_94a, Sanjuan_99}, such as losses associated with nonlinear electrical circuits. It will be shown below that the impact of this term on the behavior of the system is very similar to the impact of $b_{31}x^2 \dot x$.

The applied voltage is composed of large constant (DC) and small monochromatic components, namely, $V(t)=V_\text{DC}+v\cos\omega t$. The one dimensional equation of motion \eqref{eq:Newton_eq} can be rewritten as
\begin{multline}
    \ddot x+(2\gamma_{11}+\gamma_{31}x^2+\gamma_{32}\dot x^2)\dot x+\omega_0^2x+\alpha_3x^3\\
	=\frac{C_0\left( V_\text{DC}^2+\frac{1}{2}v^2+2V_\text{DC}v\cos\omega t+\frac{1}{2}v^2\cos 2\omega t \right)}{2md \left(1-\frac{x}{d} \right)^2}\\
	+n(t),
    \label{eq:full_eq_of_motion}
\end{multline}
where $\omega_0^2=k_1/m$, $\gamma_{11}=b_{11}/m$, $\gamma_{31}=b_{31}/m$, $\gamma_{32}=b_{32}/m$, and $\alpha_3=k_3/m$.

	\subsection{Slow envelope approximation}
	\label{sec:slow_env_approx}
In order to investigate the dynamics described by the equation of motion~\eqref{eq:full_eq_of_motion} analytically, we use the fact that nonlinearities of the micromechanical oscillator and the general energy dissipation rate are usually small (as shown in Sec.~\ref{sec:results}, the linear quality factor in our systems has a typical value of several thousands). In the spirit of the standard multiple scales method~\cite{Nayfeh_book_81, Nayfeh_Mook_book_95}, we introduce a dimensionless small parameter $\epsilon$ in~\eqref{eq:full_eq_of_motion}, and regard the linear damping coefficient $\gamma_{11}\equiv\epsilon\tilde\gamma_{11}$, the nonlinear damping coefficients $\gamma_{31}\equiv\epsilon\tilde\gamma_{31}$ and $\gamma_{32}\equiv\epsilon\tilde\gamma_{32}$, the nonlinear spring constant $\alpha_3\equiv\epsilon\tilde\alpha_3$, and the excitation amplitude $v\equiv\epsilon\tilde v$ as small. It is also assumed that the maximal amplitude of mechanical vibrations is small compared to the gap between the electrode and the mechanical beam $d$, i.e., $x/d\equiv\epsilon x/\tilde d$. Also, the frequency of excitation $\omega$ is tuned close to the fundamental mode of mechanical vibrations, namely, $\omega=\omega_0+\sigma$, where $\sigma\equiv\epsilon\tilde\sigma$ is a small detuning parameter.

Retaining terms up to first order in $\epsilon$ in~\eqref{eq:full_eq_of_motion} gives
\begin{multline}
    \ddot x+\omega_0^2x\\
	+\epsilon\left[(2\tilde\gamma_{11}+\tilde\gamma_{31}x^2+\tilde\gamma_{32}\dot x^2)\dot x+\tilde\alpha_3x^3-\frac{2}{\tilde d}x(\ddot x+\omega_0^2x)\right]\\
    =F+2\epsilon \tilde f_0\cos \omega t,
    \label{eq:motion_with_epsilon}
\end{multline}
where $F=~C_0V_{DC}^2/(2md)$, and $\epsilon\tilde f_0\equiv f_0=C_0V_{DC}v/(2md)$. We have dropped the noise from the equation of motion, and will reintroduce its averaged counterpart later in the evolution equation~\eqref{eq:evolution}.

Following \cite{Nayfeh_book_81}, we introduce two time scales $T_0=t$ and $T_1=\epsilon t$, and assume the following form for the solution:
\begin{equation*}
    x(t)=x_0(T_0,T_1)+\epsilon x_1(T_0,T_1).
\end{equation*}
It follows to the first order in $\epsilon$ that
\begin{equation*}
    \frac{d}{dt}=\frac{\partial}{\partial T_0}+\epsilon\frac{\partial}{\partial T_1},
\end{equation*}
and \eqref{eq:motion_with_epsilon} can be separated according to different orders of $\epsilon$, giving
\begin{subequations}
\begin{equation}
    \frac{\partial^2x_0}{\partial T_0^2}+\omega_0^2x_0=F,
    \label{eq:epsilon_order_0}
\end{equation}
and
\begin{multline}
    \frac{\partial^2x_1}{\partial T_0^2}+\omega_0^2x_1=2\tilde f_0\cos(\omega_0T_0+\tilde\sigma T_1)\\
    -\left(2\tilde\gamma_{11}+\tilde\gamma_{31}x_0^2+\tilde\gamma_{32}\left(\frac{\partial x_0}{\partial T_0}\right)^2\right)\frac{\partial x_0}{\partial T_0}-\tilde\alpha_3x_0^3\\
	+\frac{2F}{\tilde d}x_0-2\frac{\partial^2 x_0}{\partial T_0\partial T_1}.
    \label{eq:epsilon_order_1}
\end{multline}
\end{subequations}
The solution of \eqref{eq:epsilon_order_0} is
\begin{equation}
    x_0(T_0,T_1)=\frac{F}{\omega_0^2}+\left(a(T_1)e^{j\tilde\sigma T_1}e^{j\omega_0T_0}+\cc\right),
    \label{eq:x0}
\end{equation}
where $a$ is a complex amplitude and $\cc$ denotes complex conjugate. The "slow varying" amplitude $a$ varies on a time scale of order $T_1$ or slower.

The secular equation \cite{Nayfeh_book_81, Nayfeh_Mook_book_95}, which follows from substitution of \eqref{eq:x0} into \eqref{eq:epsilon_order_1}, is
\begin{equation}
    2\omega_0\left(j\dot a+\left(j\gamma_1-\Delta\omega\right)a\right) +\left(3\alpha_3+j\gamma_3\omega_0\right)a^2a^*=f_0,
    \label{eq:evolution_orig}
\end{equation}
where
\begin{subequations}
\begin{align}
    &\gamma_1=\gamma_{11}+\gamma_{31}\frac{F^2}{2\omega_0^4}, \label{eq:gamma_1_def}\\
    &\gamma_3=\gamma_{31}+3\omega_0^2\gamma_{32}, \label{eq:gamma_3_def}
\end{align}
\end{subequations}
and
\begin{equation*}
    \Delta\omega=\sigma-\Delta\omega_0,
\end{equation*}
where
\begin{equation}
    \Delta\omega_0=\frac{F}{\omega_0}\left(\alpha_3\frac{3F}{2\omega_0^4}-\frac{1}{d}\right)
    \label{eq:Delta_omega_0}
\end{equation}
represents a constant shift in linear resonance frequency due to the constant electrostatic force $F$. Equation \eqref{eq:evolution_orig} is also known as evolution equation. Note that we have returned to the full physical quantities, i.e, dropped the tildes, for convenience. Also, one must always bear in mind that the accuracy of the evolution equation is limited to the assumptions considered at the beginning of this Section.

As was mentioned earlier, both nonlinear dissipation terms give rise to identical terms in the evolution equation \eqref{eq:evolution_orig}. Therefore, the behavior of these two dissipation cases is similar near the fundamental resonance frequency $\omega_0$. Also, note that linear dissipation coefficient $\gamma_1$~\eqref{eq:gamma_1_def} is not constant, but is rather quadratically dependent on the constant electrostatic force $F$ due to the nonlinear dissipation term $\gamma_{31}$.

The secular equation \eqref{eq:evolution_orig} can be written as
\begin{equation}
    j\dot a +(j\gamma_1-\Delta\omega)a+q(1+jp)a^2a^*=\frac{1}{2\omega_0}\left(f_0+n_\text{slow}(t)\right),
\label{eq:evolution}
\end{equation}
where dot denotes differentiation with respect to (slow) time,
\begin{align}
    &q=\dfrac{3\alpha_3}{2\omega_0},\\
    &p=\dfrac{\gamma_3\omega_0}{3\alpha_3}, \label{eq:p}
\end{align} 
and $n_\text{slow}(t)$ is the averaged noise process with the following characteristics \cite{Dykman&Krivoglaz_84, Yurke&Buks_06}:
\begin{subequations}
	\begin{align}
		&\left<n_\text{slow}(t)\right>=0,\\
		&\left<n_\text{slow}(t)n_\text{slow}(s)\right>=N\delta(t-s),\\
		&N=\frac{k_BT}{m}\left(\gamma_1+\gamma_3|a|^2\right).
	\end{align}
\end{subequations}

\begin{figure}[htb]
    
        \includegraphics [width=3.4in]{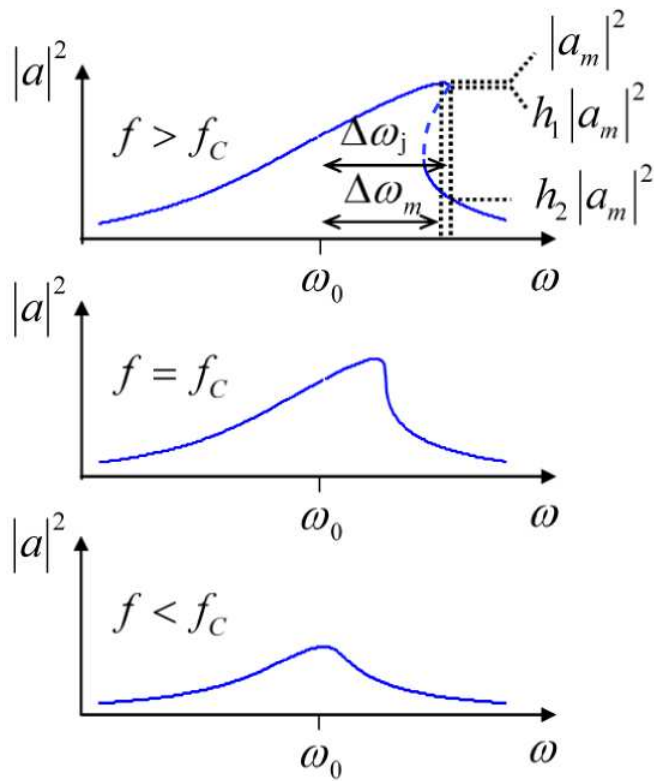}
        \caption {(Color online) Steady state solutions under different excitation amplitudes $f_0$. In case $f_0<f_C$ (where $f_C$ is some critical excitation force, dependent on the system parameters, see text), only one real solution exists and no bistability is possible. In case $f_0=f_C$, the system is on the edge of bistability and one point exists, where $|a|^2$ vs. $\omega$ has an infinite slope. In case $f_0>f_C$, the system is in bistable regime having three real solutions over some range of frequencies. Two of these solutions are stable. The dashed line denotes the unstable solution.}
        \label{fig:duffing}
    
\end{figure}

The steady state amplitude can be found by setting $\dot a=0$, $n_\text{slow}=0$ and taking a square of the evolution equation \eqref{eq:evolution}, resulting in
\begin{multline}
    q^2(1+p^2)|a|^6+2q\left(\gamma_1p-\Delta\omega\right)|a|^4\\
	+(\gamma_1^2+\Delta\omega^2)|a|^2-\frac{f_0^2}{4\omega_0^2}=0.
\label{eq:steady_ampl}
\end{multline}
This cubic equation of $|a|^2$ can have either one, two, or three different real roots, depending on the values of the detuning parameter $\Delta\omega$ and the excitation amplitude $f_0$. When $\gamma_{3}$ is sufficiently small, i.e., $p\to 0$, the solutions of \eqref{eq:steady_ampl} behave very much like the ordinary Duffing equation solutions, to which \eqref{eq:Newton_eq} reduces if $b_{31}=0$ and $b_{32}=0$ (see Fig.~\ref{fig:duffing}).%

The solution of \eqref{eq:evolution} can be also presented in polar form \cite{Nayfeh_Mook_book_95}
\begin{equation}
    a=Ae^{j\phi},
    \label{eq:a_polar}
\end{equation}
where $A$ and $\phi$ are real, and $A$ is assumed to be positive. Separating the real and imaginary parts of \eqref{eq:evolution}, one obtains (omitting the noise)
\begin{subequations}
	\begin{align}
		&\dot A+\gamma_1A+qpA^3=-\frac{f_0}{2\omega_0}\sin\phi, \label{eq:evolution_polar_sin}\\
		A&\dot \phi+\Delta\omega A-qA^3=-\frac{f_0}{2\omega_0}\cos\phi. \label{eq:evolution_polar_cos}
	\end{align}
	\label{eq:evolution_polar}
\end{subequations}
Steady state solutions are defined by $\dot A=0$, $\dot\phi=0$, which results in \eqref{eq:steady_ampl}.

The maximal amplitude $|a_m|^2$ can be found from \eqref{eq:steady_ampl} by requiring
\begin{equation*}
    \left. \frac{d(|a|^2)}{d\Delta\omega}\right|_{\Delta\omega=\Delta\omega_m}=0,
\end{equation*}
where $\Delta\omega_m$ is the corresponding excitation frequency detuning. This results in
\begin{equation}
    \frac{\Delta\omega_m}{|a_m|^2}=q=\frac{3\alpha_3}{2\omega_0}.
    \label{eq:q_vs_am}
\end{equation}

Interestingly enough, the phase $\phi$ of the maximal response is always equal $-\pi/2$, i.e., the maximal response is exactly out of phase with the excitation regardless the magnitude of the excitation, a feature well known for the linear case. This general feature can be explained as follows. For an arbitrary response amplitude $A$, there exist either two or no steady state $\phi$ solutions of \eqref{eq:evolution_polar}. If two solutions $\phi_1$ and $\phi_2$ exist, they must obey $\phi_2=\pi-\phi_1$, as seen from \eqref{eq:evolution_polar_sin}. It follows from \eqref{eq:evolution_polar_cos} that these two solutions correspond to two different values of $\Delta\omega$. However, at the point of maximum response the two solutions coincide, resulting in $\phi_1=\phi_2=-\pi/2$, i.e.,
\begin{equation}
	a_m=-j|a_m|.
	\label{eq:am_real_imag}
\end{equation}

    The system's behavior qualitatively changes when parameters such as the excitation amplitude and the frequency detuning are varied, as seen in Fig.~\ref{fig:duffing}. The parameter values at which these qualitative changes occur are called bifurcation (jump) points \cite{Strogatz_book_94}.

A jump in amplitude is characterized by the following condition:
\begin{equation*}
	\frac{d\left(|a|^2\right)}{d\Delta\omega}\to\pm\infty,
\end{equation*}
or, alternatively,
\begin{equation*}
	\frac{d\Delta\omega}{d\left(|a|^2\right)}=0.
\end{equation*}
Applying this condition to \eqref{eq:steady_ampl} yields
\begin{equation}
		3q^2(1+p^2)|a_j|^4+4q\left(\gamma_1p-\Delta\omega_j\right)|a_j|^2+(\gamma_1^2+\Delta\omega_j^2)=0,
		\label{eq:jump_point}
\end{equation}
where $\Delta\omega_j$ and $a_j$ denote the frequency detuning and the amplitude at the jump point, respectively.

When the system is on the edge of bistability, the two jump points coincide and \eqref{eq:jump_point} has a single real solution at the point of critical frequency $\Delta\omega_c$ and critical amplitude $a_c$. The driving force at the critical point is denoted in Fig.~\ref{fig:duffing} as $f_C$. This point is defined by two conditions
\begin{subequations}
    \begin{align*}
        &\frac{d\Delta\omega}{d\left(|a|^2\right)}=0,\\
        &\frac{d^2\Delta\omega}{d\left(|a|^2\right)^2}=0.
    \end{align*}
\end{subequations}
By applying these conditions one finds
\begin{equation*}
    \Delta\omega_c=\frac{3q}{2}(1+p^2)|a_c|^2+\gamma_1p,
\end{equation*}
where $a_c$ is the corresponding critical amplitude. Substituting the last result back into \eqref{eq:jump_point}, one finds  \cite{Yurke&Buks_06}
\begin{subequations}
	\begin{align}
		|a_c|^2 &=\frac{2\gamma_1}{\sqrt3q}\frac{\sqrt3p\pm1}{1-3p^2}, \label{eq:a_c}\\
		\Delta\omega_c &=\gamma_1\frac{4p\pm\sqrt3(1+p^2)}{1-3p^2},\\
		p &=\frac{\Delta\omega_c\mp\sqrt3\gamma_1}{\gamma_1\pm\sqrt3\Delta\omega_c}, \label{eq:p_from_fc}
	\end{align}
\end{subequations}
where the upper sign should be used if $\alpha_3>0$, and the lower sign otherwise.  In general, $\gamma_3$ is always positive, but $\alpha_3$ can be either positive or negative. Therefore, $q$ and $p$ are negative if $\alpha_3<0$ (soft spring), and positive if $\alpha_3>0$ (hard spring).

It follows from \eqref{eq:a_c} that the condition for the critical point to exist is
\begin{equation*}
    |p|<\frac{1}{\sqrt 3}.
\end{equation*}
Without loss of generality, we will focus on the case of "hard" spring, i.e., $\alpha_3>0$, $q>0$, $p>0$, as this is the case encountered in our experiments.

	\subsection{Behavior near bifurcation points}
	\label{sec:bifurcation_behavior}

\begin{figure} [htb]
    
        \includegraphics [width=3.4in] {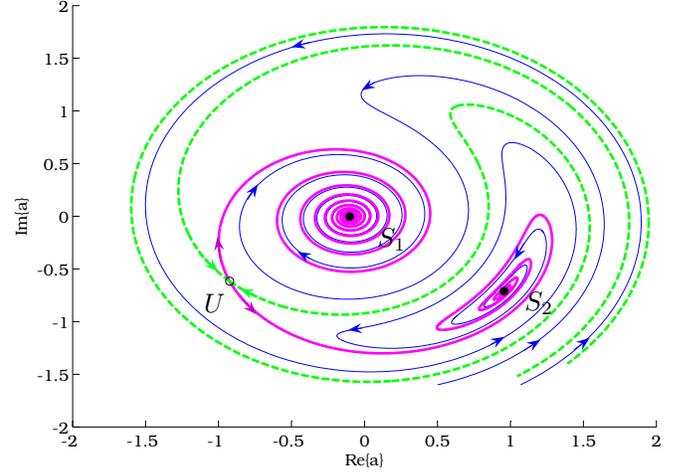}
        \caption {(Color online) Slow envelope phase plane trajectories of a nonlinear oscillator in bistable regime. Three real solutions of \eqref{eq:steady_ampl} correspond to three fixed points of the flow. $S_1$ and $S_2$ are stable spiral nodes, whereas $U$ is the saddle-node, from which two manifolds emerge \cite{Strogatz_book_94}. The green dotted line is the stable manifold ("separatrix"), which separates different basins of attraction, belonging to different stable nodes $S_1$ and $S_2$. The magenta thick line is the unstable manifold. Two typical phase plane trajectories are shown by arrowed thin blue lines.}
        \label{fig:bistable_flow}
    
\end{figure}

When the system approaches the bifurcation points, it exhibits some interesting features not existent elsewhere in the parametric phase space. In order to investigate the system's behavior in the vicinity of the jump points, it is useful to rewrite the slow envelope evolution equation \eqref{eq:evolution} as a two dimensional flow
\begin{subequations}
	\begin{align}
		\dot x&=f(x,y)+n_x(t),\\
		\dot y&=g(x,y)+n_y(t),
	\end{align}%
	\label{eq:2d_flow}
\end{subequations}%
where we have defined $x(t)=\Real\{a\}, y(t)=\Imag\{a\}$ (i.e., $a(t)=x(t)+jy(t)$), and
\begin{subequations}
	\begin{align}
		f(x,y)&=-\gamma_1x+\Delta\omega y-q(x^2+y^2)(y+px),\\
		g(x,y)&=-\Delta\omega x-\gamma_1y+q(x^2+y^2)(x-py)-\frac{f_0}{2\omega_0}.
	\end{align}
	\label{eq:2d_flow_fg}
\end{subequations}
The real-valued noise processes $n_x(t)$ and $n_y(t)$ have the following statistical properties:
\begin{subequations}
	\begin{align}
		&\left<n_c(t)\right>=\left<n_s(t)\right>=0,\\
		&\left<n_c(t)n_s(t)\right>=0,\\
		&\left<n_c(t)n_c(s)\right>=\left<n_s(t)n_s(s)\right>=\frac{N}{8\omega_0^2}\delta(t-s).
	\end{align}
	\label{eq:ncns_properties}
\end{subequations}
At the fixed points, the following holds: $f(x,y)=g(x,y)=0$. A typical phase space flow of the oscillator in bistable regime is shown in Fig.~\ref{fig:bistable_flow}.

For small displacements near an arbitrary fixed point $a_0=(x_0,y_0)$, namely, $x=x_0+\Delta x$ and $y=y_0+\Delta y$, where $\Delta x\ll x_0$ and $\Delta y\ll y_0$, the above nonlinear flow map can be approximated by its linearized counterpart
\begin{equation}
    \left(\begin{array}{c}
    \Delta\dot x \\ \Delta\dot y \end{array}\right)
    =
    M\left(\begin{array}{c}
    \Delta x \\ \Delta y\end{array}\right)
	+
	\left(\begin{array}{c}
	n_x \\ n_y\end{array}\right),
	\label{eq:linear_2d_flow}
\end{equation}
where
\begin{equation}
    M
    =
    \left(\begin{array}{cc}
    f_x & f_y \\ g_x & g_y \end{array}\right),
\end{equation}
and the excitation frequency detuning $\Delta\omega$ as well as the external excitation amplitude $f_0$ are considered constant. The subscripts in the matrix elements denote partial derivatives evaluated at $(x_0,y_0)$, for example, $$f_x\equiv\left.\frac{\partial f}{\partial x}\right|_{\begin{subarray}{l} x=x_0,\\y=y_0 \end{subarray}}.$$ The matrix $M$ is, therefore, the Jacobian matrix of the system \eqref{eq:2d_flow} evaluated at the point $(x_0,y_0)$. It is straightforward to show that
\begin{subequations}
\begin{align}
    &f_x=-\gamma_1-qp(x_0^2+y_0^2)-2qx_0(y_0+px_0),\\
    &f_y=\Delta\omega-q(x_0^2+y_0^2)-2qy_0(y_0+px_0),\\
    &g_x=-\Delta\omega+q(x_0^2+y_0^2)+2qx_0(x_0-py_0),\\
    &g_y=-\gamma_1-qp(x_0^2+y_0^2)+2qy_0(x_0-py_0).
\end{align}
\label{eq:M_elements}
\end{subequations}
Two important relations follow immediately,
\begin{subequations}
\begin{align}
    &f=g_yx-g_xy, \label{eq:gxgy} \\
    &g=f_xy-f_yx-\frac{f_0}{2\omega_0}. \label{eq:fxfy}
\end{align}
\label{eq:fxfygxgy}
\end{subequations}

The linearized system \eqref{eq:linear_2d_flow} retains the general qualitative structure of the flow near the fixed points \cite{Strogatz_book_94}, in particular both eigenvalues of the matrix $M$ are negative at the stable nodes, denoted as $S_1$ and $S_2$ in Fig.~\ref{fig:bistable_flow}. At the saddle node $U$, which is not stable, one eigenvalue of $M$ is positive, whereas the other is negative.
    
\begin{figure} [htb]
    
        \includegraphics [width=3.4in] {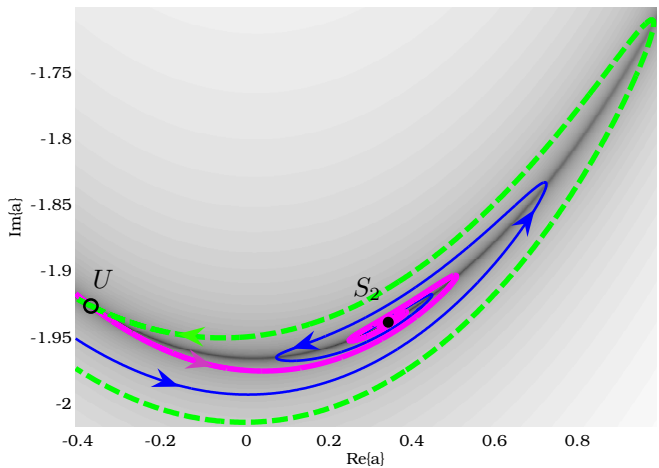}
        \caption {(Color online) The phase plane geometry when the saddle-node ($U$) and the stable node ($S_2$) are well separated. The green dotted line is the stable manifold ("separatrix") and the magenta thick line is the unstable manifold. A typical phase plane trajectory is shown by the arrowed thin blue line. The absolute value of slow envelope's rate of change $\dot{a}$ is represented by the background color. The darkest parts correspond to the slowest motion in the phase space. At both fixed points $U$ and $S_2$ the value of $\dot{a}$ is zero.}
        \label{fig:far_saddle_node}
    
\end{figure}

\begin{figure} [htb]
    
        \includegraphics [width=3.4in] {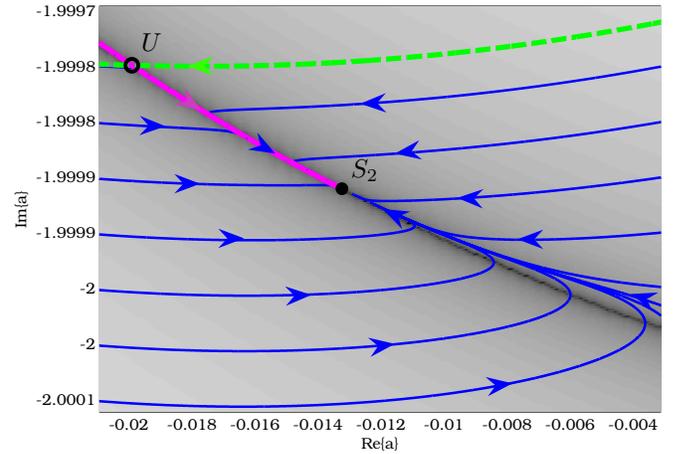}
        \caption {(Color online) The phase plane geometry when the saddle-node ($U$) and the stable node ($S_2$) are close one to another. The green dotted line is the stable manifold ("separatrix"), and the magenta thick line is the unstable manifold. Phase plane trajectories are shown by the thin blue lines. The absolute value of $\dot{a}$ is represented by the background color. The darkest parts correspond to the slowest motion in the phase space. At both fixed points $U$ and $S_2$, the value of $\dot{a}$ is zero. Note that the motion in the phase space becomes essentially one-dimensional and slows down significantly in the vicinity of the stable node $S_2$.}
        \label{fig:close_saddle_node}
    
\end{figure}

    The discussed Duffing like systems exhibit saddle point bifurcations. At the bifurcation, one of the stable nodes and the saddle node coincide, resulting in a zero eigenvalue in $M$. The bifurcation ("jump") point condition is, therefore, $\det{M}=0$, which gives the same result as in (\ref{eq:jump_point}). The case of well separated stable node and saddle node is shown in Fig.~\ref{fig:far_saddle_node}, and the case of almost coinciding stable and not stable fixed points is shown in Fig.~\ref{fig:close_saddle_node}, where the oscillator is on the verge of bifurcation.

We note that in general
\begin{equation*}
    \Tr M=f_x+g_y=-2(\gamma_1+2qp|a_0|^2),
\end{equation*}
and the slow eigenvalue near the bifurcation point can be estimated as
\begin{multline*}
    \lambda_\textrm{sd}\approx\left(\cancelto{0}{\frac{\det M}{\Tr M}}+\frac{\partial}{\partial\Delta\omega}\frac{\det M}{\Tr M}\cdot\delta\right)_{\begin{subarray}{l}\Delta\omega=\Delta\omega_j,\\ a=a_j \end{subarray}}\\
	=\frac{2q|a_j|^2-\Delta\omega_j}{\gamma_1+2qp|a_j|^2}\delta,
\end{multline*}
where $\delta$ is a small frequency detuning from $\Delta\omega_j$, i.e. $\Delta\omega=\Delta\omega_j+\delta, |\delta|\ll|\Delta\omega_j|$. If the system in bistable regime is close to bifurcation then $\lambda_\textrm{sd}\to 0$ and the evolution of the system almost comes to stagnation, phenomena often referred to as critical slowing down \cite{Yurke&Buks_06}. The motion in the vicinity of the stable node becomes slow and essentially one-dimensional along the unstable manifold. We now turn to show this analytically.

At the bifurcation points the matrix $M$ is singular , i.e., $\det M=0$. Consequently, the raws of the matrix are linearly dependent, i.e.,
\begin{equation*}
	M=
	\left(\begin{array}{cc}
        f_x & f_y\\
        rf_x & rf_y
    \end{array}\right),
\end{equation*}
where $r$ is some real constant. Using the last result, we may rewrite \eqref{eq:gxgy} at the bifurcation point as
\begin{equation*}
	r\left(f_yx-f_xy\right)=0,\\
\end{equation*}
where we have used the fact that at any fixed point (stable or saddle-node) $f(x,y)=g(x,y)=0$.
However, according to \eqref{eq:fxfy}, at any fixed point $f_yx-f_xy=-f_0/2\omega_0\neq 0$. Therefore, $r=0$ at the bifurcation point, and the matrix $M$ can be written as
\begin{equation}
	\label{eq:M_at_bifurcation}
	M=
	\lambda_f
    \left(\begin{array}{cc}
        1 & K\\
        0 & 0
    \end{array}\right),
\end{equation}
where
\begin{subequations}
	\begin{align}
		&\lambda_f=-2(\gamma_1+2qp|a_j|^2), \label{eq:lambda_f}\\
		&K=\frac{f_y}{f_x}=\frac{\gamma_1+2qp|a_j|^2}{2q|a_j|^2-\Delta\omega_j}. \label{eq:K}
	\end{align}
	\label{eq:lambda_f_and_K}
\end{subequations}
It also follows from \eqref{eq:gxgy} that
\begin{equation}
	\frac{y_j}{x_j}=\lim_{\Delta\omega\to\Delta\omega_j}\frac{g_y}{g_x}=\frac{\gamma_1+qp|a_j|^2}{\Delta\omega_j-q|a_j|^2}.
	\label{eq:yjxj}
\end{equation}

Due to the singularity of matrix $M$ at the bifurcation point, a second order Taylor expansion must be used. The flow map \eqref{eq:2d_flow} can be approximated near the bifurcation point by
\begin{subequations}
	\label{eq:2d_flow_orig_coor}
	\begin{multline}
		\Delta\dot x=\lambda_f(\Delta x + K\Delta y)+ f_\delta\delta\\
		+\frac12\left(\delta\frac{\partial}{\partial\Delta\omega}+\Delta x\frac{\partial}{\partial x}+\Delta y\frac{\partial}{\partial y}\right)^2f+n_x(t),
	\end{multline}
	\begin{multline}
		\Delta\dot y=g_\delta\delta\\
		+\frac12\left(\delta\frac{\partial}{\partial\Delta\omega}+\Delta x\frac{\partial}{\partial x}+\Delta y\frac{\partial}{\partial y}\right)^2g+n_y(t),%
	\end{multline}
\end{subequations}
where all the derivatives denoted by subscripts are evaluated at the jump point $a=a_j$, and
\begin{align*}
	f_\delta & = y_j,\\
	g_\delta & = -x_j.
\end{align*}

\begin{figure} [htb]
    
        \includegraphics [width=3.4in] {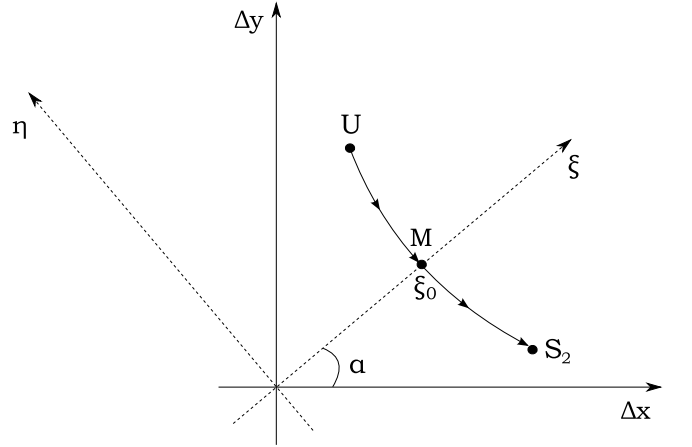}
        \caption{The effective one dimensional flow in the vicinity of a bifurcation point. The origin of the phase plane coincides with the bifurcation point. $U$ is the saddle point, and $S_2$ is a stable node. The effective flow between these two points, marked by arrows, is almost parallel to the rotated coordinate $\eta$, while the rotated coordinate $\xi$ remains essentially constant, $\xi=\xi_0+O(\eta^2)$. $\alpha$ is the angle of coordinate rotation. The velocity of the flow is largest at the point $M$, between the saddle point and the stable node.}
        \label{fig:dxdy2eta_xi}
    
\end{figure}

The above system of differential equations \eqref{eq:2d_flow_orig_coor} can be simplified by using the following rotation transformation, shown in Fig.~\ref{fig:dxdy2eta_xi},
\begin{equation}
	\left(\begin{array}{c}
        \xi\\
        \eta
    \end{array}\right)
    =
	\left(\begin{array}{cc}
        \cos\alpha & \sin\alpha\\
        -\sin\alpha & \cos\alpha
    \end{array}\right)
    \left(\begin{array}{c}
        \Delta x\\
        \Delta y
    \end{array}\right),
    \label{eq:xi_eta_definition}
\end{equation}
where $\tan\alpha=K$. 
In these new coordinates, the system \eqref{eq:2d_flow_orig_coor} becomes
\begin{subequations}
\begin{align}
	\dot\xi & = \lambda_f\xi+\Omega_\xi\delta+\frac12\D^2H(\xi,\eta)+n_\xi(t), \label{eq:xi_eq} \\ 
     \dot\eta & = -\lambda_f K\xi+\Omega_\eta\delta+\frac12\D^2E(\xi,\eta)+n_\eta(t), \label{eq:eta_eq}
\end{align}
\label{eq:xi_eta_eqs}
\end{subequations}
where
\begin{align*}
	& H(\xi,\eta) =f(\xi,\eta)\cos\alpha+ g(\xi,\eta)\sin\alpha,\\
	& E(\xi,\eta) =g(\xi,\eta)\cos\alpha- f(\xi,\eta)\sin\alpha,\\
	& \Omega_\xi =y_j\cos\alpha - x_j\sin\alpha,\\
	& \Omega_\eta =-x_j\cos\alpha - y_j\sin\alpha,\\
\end{align*}
and $\D$ is the differentiation operator
\begin{align*}
	\D & =\left(\xi\frac{\partial}{\partial\xi}+\eta\frac{\partial}{\partial\eta}+\delta\frac{\partial}{\partial\Delta\omega}\right)\\
	& =\left(\Delta x\frac{\partial}{\partial x}+\Delta y\frac{\partial}{\partial y}+\delta\frac{\partial}{\partial\Delta\omega}\right).\\
\end{align*}
The noise processes $n_\xi(t)$ and $n_\eta(t)$ have the same statistical properties \eqref{eq:ncns_properties} as $n_c(t)$ and $n_s(t)$.

The time evolution of the system described by the differential equations \eqref{eq:xi_eta_eqs} has two distinct time scales. Motion along the coordinate $\xi$ is "fast", and settling time is of order $|\lambda_f|^{-1}$.  The time development along the coordinate $\eta$, however, is much slower, as will be shown below.

On a time scale much longer than $|\lambda_f|^{-1}$, the coordinate $\xi$ can be regarded as not explicitly dependent on time. The momentary value of $\xi$ can be approximated as
\begin{equation}
    \xi = -\frac{1}{\lambda_f}\left(\Omega_\xi \delta+\frac{1}{2}\frac{\partial^2H}{\partial \eta^2}\eta^2\right),
	\label{eq:xi_approx}
\end{equation}
where we have neglected all terms proportional to $\delta^2$ and $\delta\eta$.

The motion along the coordinate $\eta$ is governed by a slow evolution equation \eqref{eq:eta_eq}, combining which with \eqref{eq:xi_approx} results in
\begin{equation}
     \dot\eta=\dot\eta_0+B\eta^2+n_\eta(t),
     \label{eq:one_dim_flow}
\end{equation}
where
\begin{subequations}
	\begin{equation}
		\dot\eta_0 = -\frac{x_j}{\cos\alpha}\delta,
		\label{eq:eta_0}
	\end{equation}
	\begin{multline}
		B = \frac{q}{\cos\alpha} \big[x_j(1+2\sin^2\alpha+p\sin 2\alpha) \\
		-y_j(p(1+2\cos^2\alpha)+\sin 2\alpha) \big].
		\label{eq:B}
	\end{multline}
	\label{eq:eta_0_and_B}
\end{subequations}
Note that the noise process $n_\xi(t)$ does not play a significant role in the dynamics of the system, because the system is strongly confined in $\xi$ direction. Such noise squeezing is a general feature of systems nearing saddle-point bifurcation \cite{Yurke&Buks_06, Almog_et_al_07, Yurke_et_al_95, Rugar&Grutter_91}. 

\begin{figure} [htb]
    
        \includegraphics [width=3.4in] {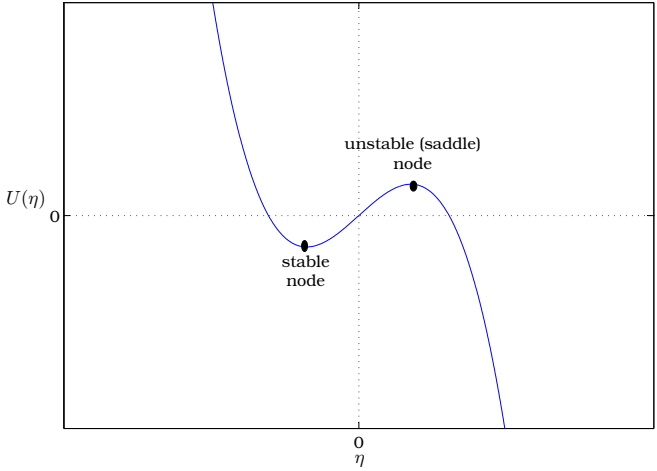}
        \caption{(Color online) Effective one dimensional potential $U(\eta) \propto -\eta\left(\dot\eta_0+\frac13B\eta^2\right)$ \eqref{eq:one_dim_potential}.}
        \label{fig:one_dim_potential}
    
\end{figure}

\begin{figure} [htb]
    
        \includegraphics [width=3.4in] {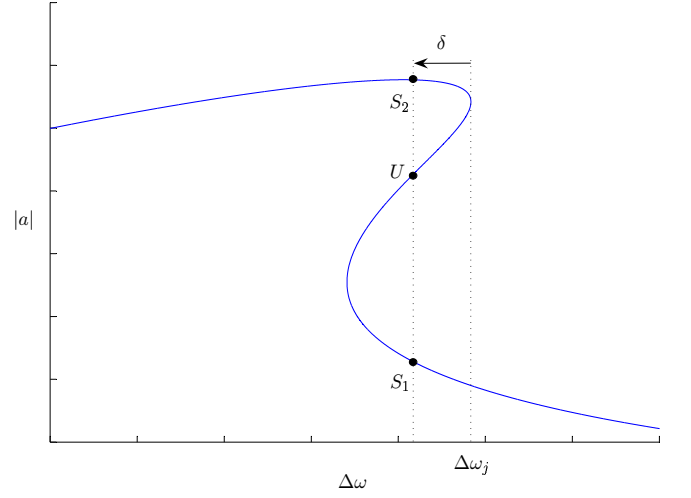}
        \caption{(Color online) The location of stable nodes $S_1$ and $S_2$, and a saddle node $U$ in a bistable regime close to a bifurcation point. $\delta$, which is negative in this case, is the frequency difference between the excitation frequency and the jump point frequency $(\omega_0+\Delta\omega_0)+\Delta\omega_j$. The scales of the axes are arbitrary.}
        \label{fig:one_dim_simulation_fdomain}
    
\end{figure}

\begin{figure} [htb]
    
        \includegraphics [width=3.4in] {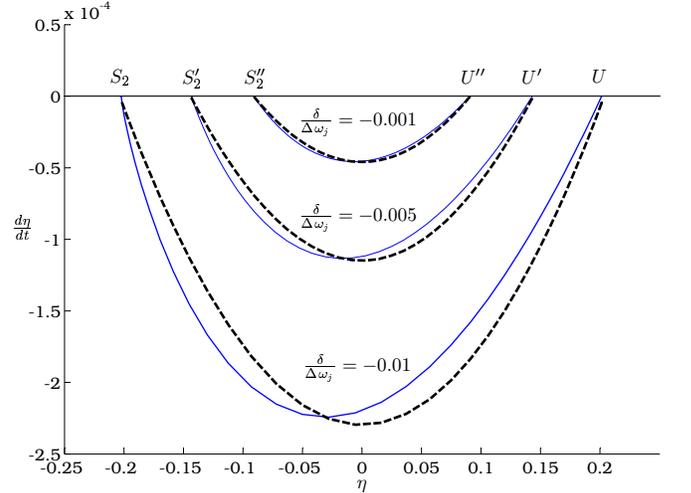}
        \caption{(Color online) Velocity along the slow coordinate $\eta$ for different values of detuning $\delta$. $\Delta\omega_j$ is the jump point (bifurcation) detuning. $p=0.3/\sqrt{3}$ in all cases, $U, U'$ and $U''$ are the saddle node positions for different values of $\delta$. Similarly, $S_2, S_2'$ and $S_2''$ are the stable node positions for different values of $\delta$. Exact values of $d\eta/dt$ are shown by solid lines. The dashed lines are the results of analytical approximation \eqref{eq:one_dim_flow}. The scales of the axes are arbitrary.}
        \label{fig:one_dim_simulation}
    
\end{figure}

Two qualitatively different cases of \eqref{eq:one_dim_flow} should be recognized. The first case is of a system in a bistable regime with a stable (quasi stable, as we will see below) and non stable (saddle) fixed points close enough to a bifurcation point. In this case, the one dimensional motion is equivalent to a motion of a massless particle in a confining cubic "potential"
\begin{equation}
	U(\eta) = -\eta \left(\dot\eta_0+\frac13B\eta^2 \right),
	\label{eq:one_dim_potential}
\end{equation}
as shown in Fig.~\ref{fig:one_dim_potential}.
Figure~\ref{fig:one_dim_simulation_fdomain} depicts the location of the fixed points and the bifurcation point on a frequency response curve in this case. Figure~\ref{fig:one_dim_simulation} shows a comparison between the exact simulation of the system's motion near the bifurcation point and the analytical result \eqref{eq:one_dim_flow}.

The quasi one-dimensional system described above is obviously not stable \cite{Kramers_40, Chan_et_al_08}. The rate of escape from the vicinity of the quasi stable fixed point is \cite{Kramers_40, Hanggi_et_al_90}
\begin{equation*}
	r_\text{therm}(\delta ) \approx \frac{\omega_A\omega_B}{2\pi}e^{-\frac{\Delta U}{D}},
\end{equation*}
where
\begin{align*}
	\Delta U&=\frac{4}{3}\sqrt{-\frac{\dot\eta_0^3}{B}},\\
	D&=\frac{N}{16\omega_0^2}=\frac{k_BT}{16m\omega_0^2}\left(\gamma_1+\gamma_3|a_j|^2\right),\\
	\omega^2_A&=\left.\frac{\partial^2U}{\partial \eta^2}\right|_{\text{stable node}}=2\sqrt{-\dot\eta_0 B},\\
	\omega_B&=-\omega_A.
\end{align*}
Characteristic time of thermal escape $\tau_\text{therm}$ can be shown to be \cite{Dykman&Krivoglaz_84}
\begin{subequations}
\begin{equation}
	\tau_\text{therm}(\delta )=\frac{1}{r_\text{therm}} \approx \tau_0 e^{\frac{\Delta}{k_BT}},
\end{equation}
where
\begin{align}
	\tau_0&=\frac{\pi}{\sqrt{-\dot\eta_0 B}} \propto (-\delta)^{-\frac{1}{2}},\\
	\Delta&=\frac{64}{3}\frac{m\omega_0^2}{\gamma_1+\gamma_3|a_j|^2}\sqrt{\frac{-\dot\eta_0^3}{B}} \propto (-\delta)^\frac{3}{2}.
\end{align}
\label{eq:tau_therm}
\end{subequations}
This is a mean time in which the system escapes from the stable node near bifurcation point to the other stable solution of \eqref{eq:evolution} due to thermal noise $n_\eta(t)$, and the $3/2$ power law is correct as long as $\Delta\gg k_BT$ \cite{Dykman_et_al_04}.

The second case describes a system which has undergone saddle bifurcation, i.e., an annihilation of the stable and non stable points has occurred. The phase plane motion close to the bifurcation point is still one dimensional, however, $\dot\eta_0$ changes its sign. Therefore, the motion is not confined any more, but is still very slow in the vicinity of the bifurcation point, because $\dot\eta_0\propto\delta$, as follows from Eq.~\eqref{eq:eta_0}. The system starts converging to the single remaining stable  fixed point, but is significantly slowed down, and lingers in the vicinity of the bifurcation point due to the saddle node "ghost". As the system spends most of its time of convergence near the saddle node "ghost", this slow time of convergence $\tau_\text{sd}$ can be roughly estimated as \cite{Strogatz_book_94}
\begin{equation}
	\tau_\text{sd}=\int_0^\infty \frac{d\eta}{\dot\eta_0+B\eta^2}=\frac{\pi}{2\sqrt{\dot\eta_0 B}}.
	\label{eq:tau_sd}
\end{equation}
Note that $\tau_\text{sd}\propto\delta^{-\frac{1}{2}}$, due to \eqref{eq:eta_0}.

	\subsection{Extraction of parameters from experimental data}
	\label{sec:parameter_extraction}
	
The analytical results presented above allow us to use data acquired in relatively simple experiments in order to estimate several important dynamic parameters of the micromechanical beam. We note that data acquisition using e-beam or optical beam interaction with vibrating elastic element does not readily enable extraction of displacement values. In contrast, the frequencies of important dynamical features, including maximum and jump points, can be measured with high accuracy using standard laboratory equipment, such as network analyzers and lockin amplifiers. Therefore, it is desirable to be able to extract as much data as possible from the frequency measurements.

If the system can be brought to the verge of bistable regime, i.e., $f_0=f_c$, the nonlinear damping parameter $p$ can be readily determined using Eq.~\eqref{eq:p_from_fc}. The same coefficient can also be extracted from the measurements of the oscillator's frequency response in the bistable regime. In general, the sum of the three solutions for $|a|^2$ at any given frequency can be found from Eq.~\eqref{eq:steady_ampl}. This is employed for the jump point at $\omega_{0}+\Delta\omega_0+\Delta\omega_j$ seen in Fig. \ref{fig:duffing}. Using Eq.~\eqref{eq:q_vs_am} to calibrate the measured response at this jump point one has
\begin{equation*}
	(2h_1+h_2)|a_m|^2=-\frac{2q\left(\gamma_1p-\Delta\omega_j\right)}{q^2(1+p^2)},
\end{equation*} or
\begin{equation}
    \left(2h_1+h_2\right)\Delta\omega_m(1+p^2)+2\left(\gamma_1p-\Delta\omega_j\right)=0,
    \label{eq:p_quadratic_eq}
\end{equation}
where $h_1$ and $h_2$ are defined in Fig.~\ref{fig:duffing}. Due to the frequency proximity between the maximum point and the jump point at $\omega=\omega_0+\Delta\omega_0+\Delta\omega_j$, the inaccuracy of such a calibration is small. Moreover, as long as excitation amplitude is high enough, $h_2$ is much smaller than $h_1$ and even considerable inaccuracy in $h_{2}$ estimation will not have any significant impact. This equation can be used to estimate $p$ for different excitation amplitudes at which the micromechanical oscillator exhibits bistable behavior, i.e., $f_0>f_c$. It is especially useful if the system is strongly nonlinear and cannot be measured near its critical point due to high noise floor or low sensitivity of the displacement detectors used.

Another method for estimating the value of $p$ requires measurement of free ring down transient of the micromechanical oscillator and can be employed also at low excitations, when the system does not exhibit bistable behavior, i.e., $f_0<f_c$. The polar form of the evolution equation \eqref{eq:evolution_polar} is especially well suited for the analysis of  the system's behavior in time domain. Starting from Eq.~\eqref{eq:evolution_polar_sin} and applying the free ring down condition $f_0=0$, one finds
\begin{equation}
	A^2(t) =\frac{A_0^2e^{-2\gamma_1t}}{1+\frac{qp}{\gamma_1}A_0^2(1-e^{-2\gamma_1t})},
	\label{eq:A^2_ringdown}
\end{equation}
where $A_0$ is the amplitude at $t=0$. In particular, consider a case in which the system is excited at its maximal response frequency detuning $\Delta\omega_m$, i.e., $A_0^2=|a_m|^2$. Then, after turning the excitation off,  the amplitude during the free ring down process described by Eq.~\eqref{eq:A^2_ringdown} can be written as
\begin{equation}
	\frac{A^2(t)}{|a_m|^2} =\frac{e^{-2\gamma_1t}}{1+p\frac{\Delta\omega_m}{\gamma_1}(1-e^{-2\gamma_1t})}.
	\label{eq:A_m^2_ringdown}
\end{equation}
The ring down amplitude measured in time domain can be fitted to the last result.

In addition to nonlinear damping parameter $p$, most parameters defined above can be easily estimated from frequency measurements near the jump point shown in Fig.~\ref{fig:duffing} if the following conditions are satisfied. The first condition is
\begin{subequations}
\begin{equation}
	\left| \frac{\Delta\omega_j-\Delta\omega_m}{\Delta\omega_j} \right| \ll 1,
\end{equation}
which can be satisfied by exciting the  micromechanical beam oscillator in the bistable regime strongly enough, i.e., $f_0 \gg f_c$. The immediate consequence of the first condition is
\begin{equation}
	\frac{|a_m|^2-|a_j|^2}{|a_j|^2} \ll 1,
	\label{eq:am_approx_aj}
\end{equation}
\end{subequations}
i.e., $h_1 \approx 1$, as described above.

Using Eq.~\eqref{eq:am_approx_aj}, it follows from Eq.~\eqref{eq:q_vs_am} that $q|a_j|^2 \approx \Delta\omega_m$. From the last result and from Eqs.~\eqref{eq:lambda_f_and_K}, \eqref{eq:yjxj}, and \eqref{eq:xi_eta_definition}, the following approximations follow immediately:
\begin{subequations}
	\begin{align}
		K &\equiv \tan\alpha \approx \frac{\gamma_1}{\Delta\omega_m}+2p, \label{eq:approx_K} \\
		\lambda_\text{sd} &=\frac{1}{K}\delta, \\
		\lambda_f &\approx -2(\gamma_1+2p\Delta\omega_m),\\
		\frac{y_j}{x_j}&\approx \frac{\gamma_1+p\Delta\omega_m}{\Delta\omega_j-\Delta\omega_m}.
	\end{align}
	\label{eq:approx_K_lf_yjxj}
\end{subequations}

As shown in Sec.~\ref{sec:slow_env_approx}, Eq.~\eqref{eq:am_real_imag}, at the maximum response point $\Delta\omega=\Delta\omega_m$, the following holds: $a_m=-j|a_m|$. Therefore, in view of our assumptions described above, we may write
\begin{equation*}
	x_j \approx -|a_j|\left(\frac{y_j}{x_j}\right)^{-1}.
\end{equation*}
Consequently,
\begin{subequations}
	\begin{equation}
			\dot\eta_0 \approx |a_j|\left(\frac{y_j}{x_j}\right)^{-1}\sqrt{1+K^2}\delta,
	\end{equation}
	and
	\begin{multline}
			B \approx \frac{\Delta\omega_m}{|a_j|(1+K^2)^\frac32}\left(\frac{y_j}{x_j}\right)^{-1}\biggl[ 2K\left(\frac{y_j}{x_j}\right)-3K^2-1\\
	+p\left((3+K^2)\left(\frac{y_j}{x_j}\right)-2K\right) \biggl],
	\end{multline}
\end{subequations}
which follows from Eq.~\eqref{eq:eta_0_and_B}. The time $\tau_\text{sd}$, which describes the slowing down near the saddle-node "ghost" described above Eq.~\eqref{eq:tau_sd}, can be expressed as
\begin{widetext}
\begin{equation}
	\tau_\text{sd}(\delta) = \frac{\pi Y}{2\sqrt{\delta}},
	\label{eq:tau_sd_approx}
\end{equation}
where
\begin{equation}
	Y\equiv\sqrt{\frac{\delta}{B\dot\eta_0}}\approx \frac{\left(\frac{y_j}{x_j}\right)(1+K^2)}{\sqrt{\Delta\omega_m \left[2K\left(\frac{y_j}{x_j}\right)-3K^2-1+p\left((3+K^2)\left(\frac{y_j}{x_j}\right)-2K\right)\right]}}.
	\label{eq:approx_Y}
\end{equation}
\end{widetext}

Finally, we turn to estimate the value of the thermal escape time $\tau_\text{therm}$ given by Eq.~\eqref{eq:tau_therm}. Using the same assumptions as above, we find
\begin{subequations}
	\begin{align}
		\tau_0 &\approx \frac{\pi Y}{\sqrt{-\delta}}, \\
		\Delta U &\approx \frac{64}{3} \frac{m\omega_0^2}{\gamma_1+\gamma_3|a_m|^2} \left(\frac{y_j}{x_j}\right)^{-2}|a_m|^2\sqrt{1+K^2}Y\left(-\delta\right)^{\frac32}.
	\end{align}
	\label{eq:tau_therm_approx}
\end{subequations}
Unlike in the previous approximations, one has to know at least the order of magnitude of the response amplitude in the vicinity of the jump point (in addition to effective noise temperature $T$ and effective mass $m$) in order to approximate $\tau_\text{therm}$ appropriately. The same is also true for estimation attempt of the physical nonlinear constants
\begin{subequations}
	\begin{align}
		\alpha_3 &=\frac{2\omega_0\Delta\omega_m}{3|a_m|^2}, \\
		\gamma_3 &=2p\frac{\Delta\omega_m}{|a_m|^2}.
	\end{align}
	\label{eq:alpha3_gamma3_approx}
\end{subequations}
For more accurate estimation, one of several existing kinds of fitting procedures can be utilized \cite{Dick_et_al_06, Jaksic&Boltezar_02}.
However, the order of magnitude estimations often fully satisfy the practical requirements.

	\subsection{Experimental considerations}
	\label{sec:sweep_cond}

The above discussion of parameters' evaluation using experimental data, especially in frequency domain, emphasizes the importance of accurate frequency measurements. However, the slowing down of the oscillator's response near the bifurcation points poses strict limitations on the rates of excitation frequency or amplitude sweeps used in such measurements \cite{Kogan_07}. This is to say that special care must be taken by the experimentalist choosing a correct sweep rate for the measurement in order to obtain the smallest error possible. Fortunately, this error can easily be estimated based on our previous analysis.

Let $r_\text{sweep}$ represent the frequency sweep rate in the frequency response measurement. For example, using network analyzer in part of our experiments, we define
\begin{equation}
	r_\text{sweep}=2\pi\frac{\text{frequency span } (\hertz)}{\text{sweep time }(\sec)}.
	\label{eq:r_sweep}
\end{equation}
In order to estimate the inaccuracy, $\delta_\text{err}$, in the measured value of the bifurcation point detuning, $\Delta\omega_j$, which results from nonzero frequency sweep rate, the following expression may be used:
\begin{equation*}
	\frac{|\delta_\text{err}|}{r_\text{sweep}} \approx \tau_\text{sd}(\delta_\text{err}),
\end{equation*}
whose solution is
\begin{equation}
	\delta_\text{err} \approx \left(\frac{\pi}{2}Y r_\text{sweep}\right)^{\frac23}.
	\label{eq:delta_err}
\end{equation}
Note that this error is a systematic one - the measured jump point will always be shifted in the direction of the frequency sweep. Obviously, the first step towards accurate measuring of $\Delta\omega_j$ is to ensure that the established value of the bifurcation point detuning does not change when the sweep rate is further reduced.

Another possible source of uncertainty in frequency measurements near the bifurcation point is the thermal escape process. The error introduced by this process tends to shift the measured jump point detuning in the direction opposite to the direction of the frequency sweep. Moreover, unlike the error arising from slowing down process, this inaccuracy cannot be totally eliminated by reducing the sweep rate. However, as will be shown in Sec.~\ref{sec:results_parameter_extraction}, in our case this error is negligible.

\section{Results}
\label{sec:results}

	\subsection{Nonlinear damping}
	\label{sec:results_nonlin_damping}
\begin{figure}[htb]
    
        \includegraphics[width=3.4in]{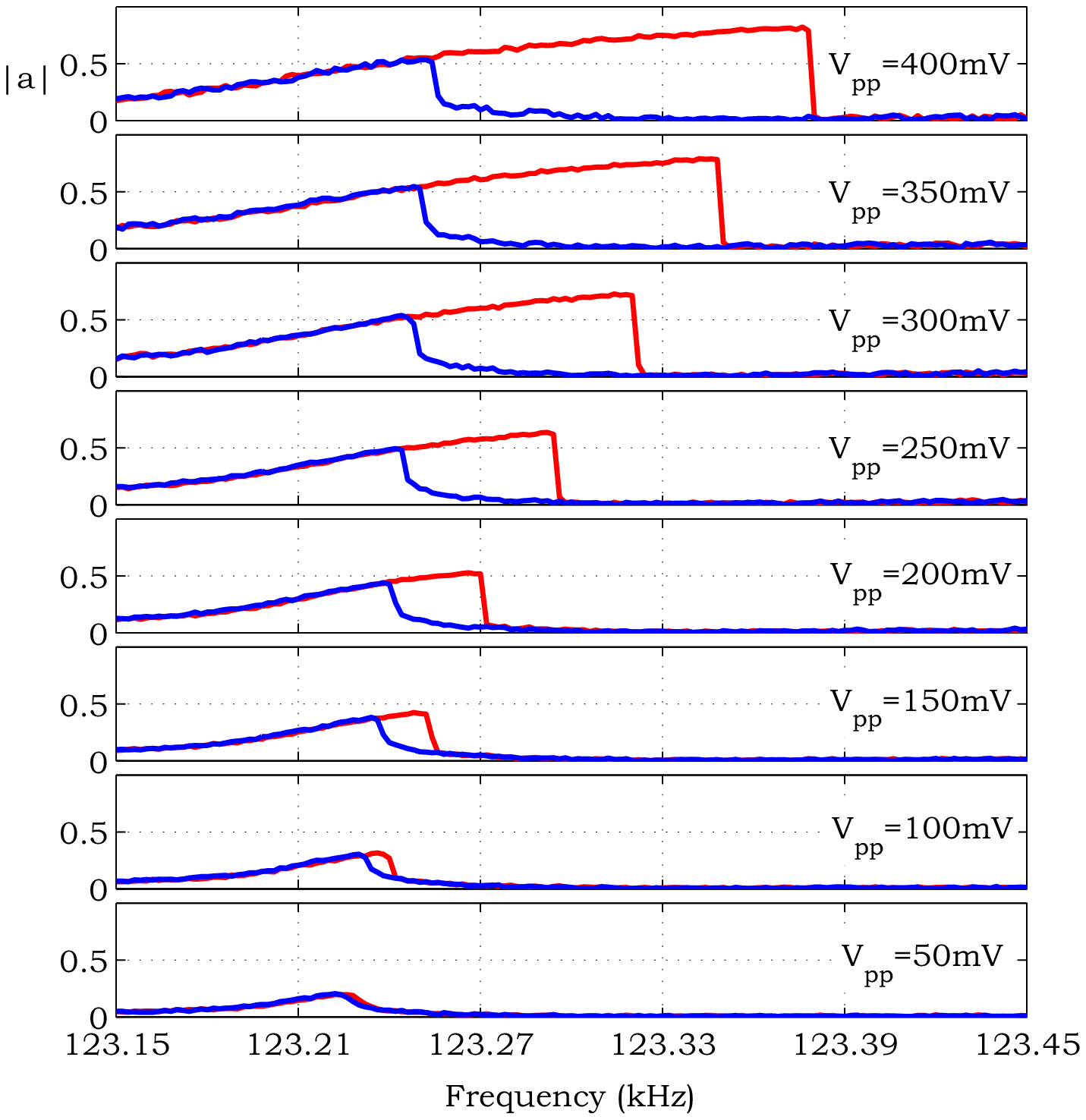}
        \caption{(Color online) Measured response amplitude vs. excitation frequency shown for both upward and downward frequency sweeps with $V_{DC}=20\volt$ and with varying peak-to-peak excitation amplitude $v_{ac}$ of a 200\micm long beam with fundamental mode occurring at 123.2\kHz. The excitation amplitude is shown on the graphs. The oscillator exhibits bistable behavior at all excitation amplitudes except for the lowest one. The vertical axis is in arbitrary units.}
        \label{fig:raw_data}
    
\end{figure}

\begin{figure}[htb]
    
        \includegraphics[width=3.4in]{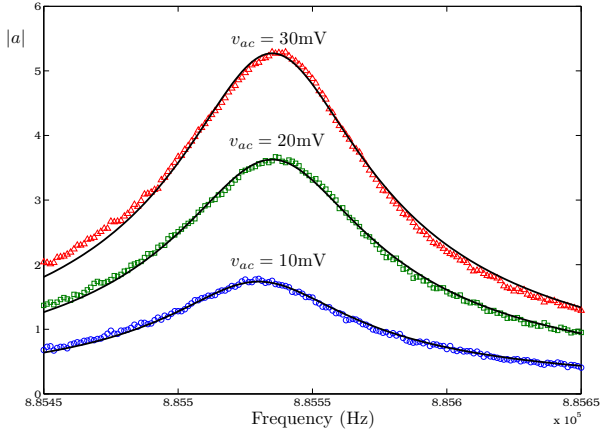}
        \caption{(Color online) Measured response vs. frequency in a linear regime of the 125\micm long beam with fundamental mode occurring at 885.53\kHz and $V_{DC}=15\volt$. The linear regime is defined as a regime in which the frequency response function is symmetric around the resonance frequency. In the main panel, the measured responses with three different excitation amplitudes are shown.  Blue circles correspond to $v=10\millivolt$, green rectangles correspond to $v=20\millivolt$, and red triangles correspond to $v=30\millivolt$. Solid black lines show the fitted Lorentzian shapes. Vertical scale is in arbitrary units.}
        \label{fig:experiment_linear_NA}
    
\end{figure}

\begin{figure}[htb]
    
        \includegraphics[width=3.4in]{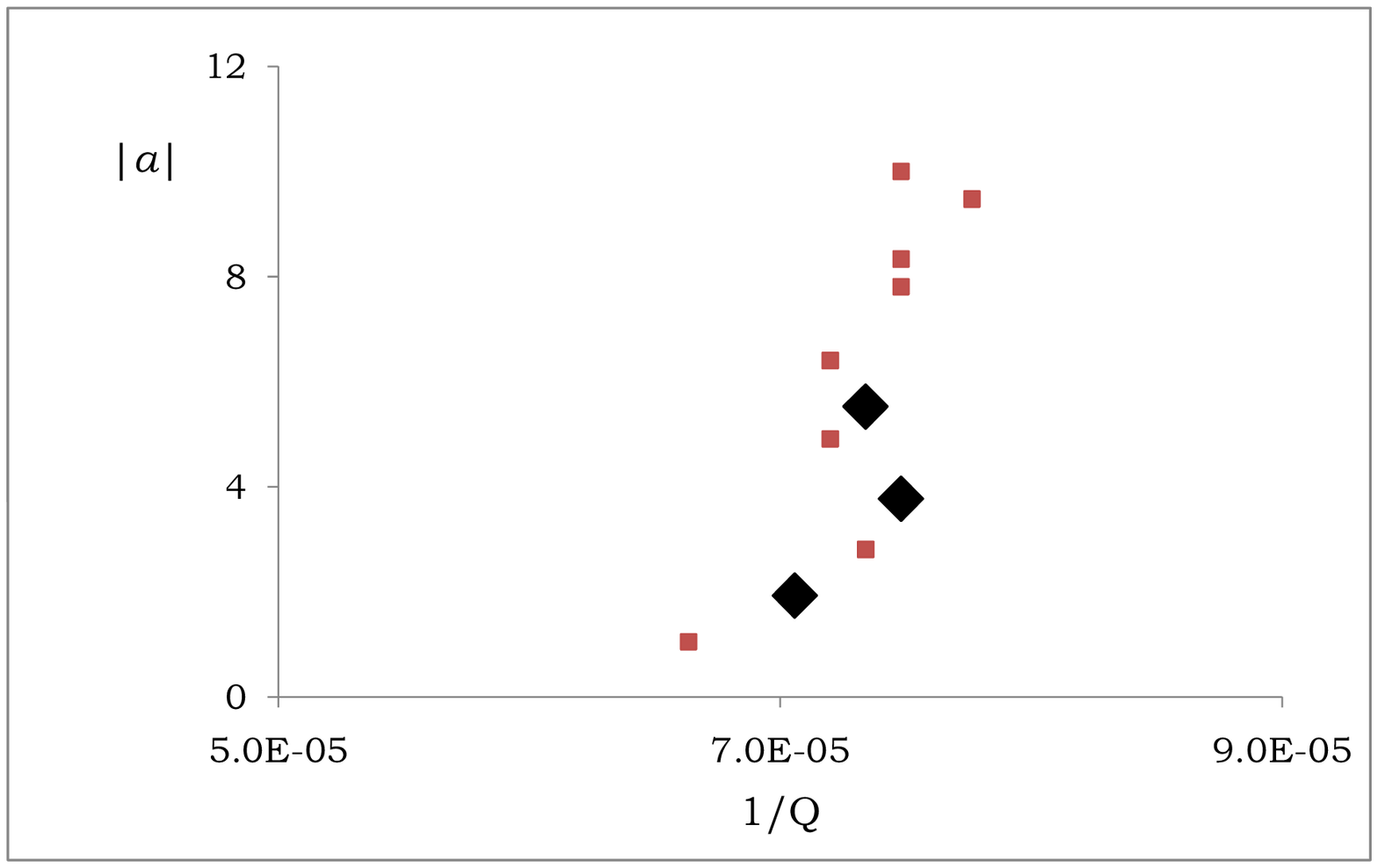}
        \caption{(Color online) Measured response amplitude $|a|$ vs. inverse quality factor $1/Q$ in a linear regime \cite{Gottlieb&Feldman_97, Husain_et_al_03}. Large black diamonds correspond to the frequency responses depicted in Fig.~\ref{fig:experiment_linear_NA}. The measured averaged quality factor is $Q=13600\pm 4\%$. Other experimental parameters are similar to those described in Fig.~\ref{fig:experiment_linear_NA} caption. Vertical scale is in arbitrary units.}
        \label{fig:quality_factor_backbone}
    
\end{figure}

A typical measured response of the fundamental mode of a 200\micm long beam occurring at the resonance frequency of 123.2\kHz measured with $V_{DC}=20\volt$ and varying excitation amplitude is seen in Fig.~\ref{fig:raw_data}. The linear regime is shown in the frequency response diagram and damping backbone curve depicted in Figs.~\ref{fig:experiment_linear_NA} and \ref{fig:quality_factor_backbone} respectively, for a 125\micm long beam with fundamental mode resonant frequency 885.53\kHz and $V_{DC}=20\volt$. We derive the value of $\gamma_1=\omega_0/2Q$ from the linear response at low excitation amplitude and find $Q=7200$ for 200\micm beam and $Q=13600$ for 125\micm beam.

\begin{figure}[htb]
    
        \includegraphics[width=3.4in]{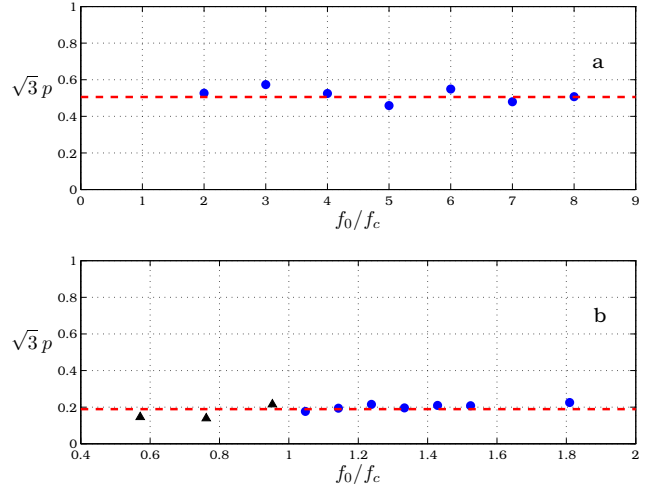}
        \caption{(Color online) Experimental results for $p=\gamma_3\omega_0/3\alpha_3$ vs.~excitation amplitude. The excitation amplitude on the horizontal axis is normalized by the respective critical excitation amplitude $f_c$.
\textbf{a.}~200\micm long beam with fundamental mode occurring at 123.20\kHz and $Q=7200$. The values of $p$ extracted from frequency domain jump point measurements (see Eq.~\eqref{eq:p_quadratic_eq}) are represented by blue circles. Red dashed line represents the value of $p=0.505/\sqrt{3}=0.292$ evaluated using the critical point frequency detuning $\Delta\omega_c$ (see Eq.~\eqref{eq:p_from_fc}). The critical excitation voltage is 50\millivolt, and $V_{DC}=20\volt$.
\textbf{b.}~120\micm long beam with fundamental mode occurring at 885.53\kHz and $Q=13600$. The values of $p$ extracted from time domain ring down measurements according to Eq.~\eqref{eq:A_m^2_ringdown} and frequency domain jump point measurements (see Eq.~\eqref{eq:p_quadratic_eq}) are represented by green squares and blue circles respectively. Red dashed line represents the value of $p=0.189/\sqrt{3}=0.109$ evaluated using the critical point frequency detuning $\Delta\omega_c$ that is given by Eq.~\eqref{eq:p_from_fc}. The critical excitation voltage is 105\millivolt, and $V_{DC}=15\volt$.}
        \label{fig:experiment_p_extract}
    
\end{figure}

As shown in Sec.~\ref{sec:parameter_extraction}, the value of $p$ can be estimated for different excitation amplitudes using Eqs.~\eqref{eq:p_quadratic_eq} and \eqref{eq:A_m^2_ringdown}. Typical results of applying these methods to experimental data from a micromechanical beam oscillator can be seen in Fig.~\ref{fig:experiment_p_extract}. Using these procedures we find $p\approx 0.292$ for the 200\micm long beam and $p\approx 0.109$ for the 125\micm long beam. We also estimate $p$ from the critical point detuning using Eq.~\eqref{eq:p_from_fc}, and obtain similar values.

	\subsection{Parameter evaluation}
	\label{sec:results_parameter_extraction}

In order to illustrate the procedures derived in Sec.~\ref{sec:parameter_extraction}, we evaluate the main parameters of slow envelope dynamics of the 125\micm long beam in a particular case in which $V_{DC}=15\volt$ and the excitation voltage amplitude is 140\millivolt. The quality factor of the beam, as measured in the linear regime, is $Q=13600$.

The results that can be derived from frequency measurements only, i.e., the results corresponding to Eqs.~\eqref{eq:p_from_fc}, \eqref{eq:approx_K_lf_yjxj} and \eqref{eq:approx_Y}, are summarized in Table~\ref{tab:eval_freq_meas_parameters}.
\begin{table}[htb!]
	\centering
	\begin{tabular}{l  c  c}
		\hline
		Parameter & Value & Units \\[0.5ex]
		\hline
		$\omega_0/2\pi$ & $885534$ & $\hertz$ \\
		$\gamma_1$ & $204$ & $\sec^{-1}$ \\
		$\Delta\omega_m/2\pi$ & $76$ & $\hertz$ \\
		$\Delta\omega_j/2\pi$ & $81$ & $\hertz$ \\
		$p$ & 0.109 & \\
		$K$ & 0.646 & \\
		$\alpha$ & 0.573 & $\radian$ \\
		$\lambda_f$ & -616.5 & $\sec^{-1}$\\[0.5ex]
		$\frac{y_j}{x_j}$ & 8.16 & \\[0.5ex]
		$Y$ & 0.158 & $\sec^{-\frac12}$ \\[1ex] 	
		\hline
	\end{tabular}
	\caption{Parameters of the slow envelope dynamics of a 125\micm long beam. Applied DC voltage is 15\volt and excitation voltage amplitude is 140\millivolt. The critical excitation voltage is 105\millivolt. Quality factor is $Q=13600$.}
	\label{tab:eval_freq_meas_parameters}
\end{table}

For this measurement we employ a network analyzer with frequency span of 500\hertz, sweep time of $13.6\sec$, and bandwidth of 18\hertz. Therefore, the sweep rate defined in Eq.~\eqref{eq:r_sweep} is
\begin{equation*}
	r_\text{sweep}=2\pi\frac{500\hertz}{13.6\sec}=231\radian\sec^{-2}.
\end{equation*}
The inaccuracy in jump point detuning estimation due to slowing down process (see Eq.~\eqref{eq:delta_err}) is
\begin{equation}
	\frac{\delta_\text{err}}{2\pi} \approx 2 \hertz.
	\label{eq:delta_err_value}
\end{equation}

We now turn to estimate the order of magnitude of other parameters, including the nonlinear elastic constant $\alpha_3$ and nonlinear damping constant $\gamma_3$. Based on the observations of the vibrating micromechanical beam by the means of SEM continuous scanning mode, we estimate the amplitude of mechanical vibration to be around 100\nanometer. The mass of a golden beam of the dimensions given in Sec.~\ref{sec:exp_setup} is approximately $7\times10^{-13}\kg$. These estimations allow us to assess the order of magnitude of several additional parameters shown in Table~\ref{tab:eval_special_parameters}, which is based on Eqs.~\eqref{eq:tau_therm_approx} and \eqref{eq:alpha3_gamma3_approx}.
\begin{table}[htb]
	\centering
	\begin{tabular}{l  c  c}
		\hline
		Parameter & Value at & Units \\
		& $\delta=-\delta_\text{err}=-2\pi\times 2\hertz$ &\\[0.5ex]
		\hline
		$\alpha_3$ & $2\times 10^{23}$ & $\meter^{-2}\sec^{-2}$ \\
		$\gamma_3$ & $1\times 10^{16}$ & $\meter^{-2}\sec^{-1}$ \\
		$T$ & 300 & \Kelvin\\
		$\Delta U/k_BT$ & $6\times 10^5$ & \\
		$\tau_0$ & $0.13$ & $\sec$ \\[1ex] 	
		\hline
	\end{tabular}
	\caption{Order of magnitude estimation of parameters of a 125\micm long beam's slow envelope dynamics. The distance from the excitation frequency to the jump frequency is taken to be equal to $\delta_\text{err}$ (see Eq.~\eqref{eq:delta_err_value}). Applied DC voltage is 15\volt, the excitation voltage amplitude is 140\millivolt, and the estimated amplitude of vibration is 100\nanometer. The critical excitation voltage is 105\millivolt. Quality factor is $Q=13600$.}
	\label{tab:eval_special_parameters}
\end{table}

We estimate below the thermal escape time for $\delta=-\delta_\text{err}$ (see Eq.~\eqref{eq:delta_err_value}). However, the value of the exponent, $\Delta U/k_BT \sim 6\times 10^5$ at $T=300\kelvin$, makes the thermal escape time at this detuning value extremely large. Therefore, in our experiments, the thermal escape process does not contribute significantly to the total inaccuracy in frequency measurements near the bifurcation point, at least for effective noise temperatures lower than $10^8\kelvin$, at which the assumption $\Delta\gg k_BT$ is no longer valid.

Finally, it is interesting to compare the nonlinear dissipation term $\gamma_3|a|^2$ and the linear dissipation term $\gamma_1$ in the evolution equation~\eqref{eq:evolution}. It follows from the above assumptions and the values in Table~\ref{tab:eval_special_parameters} that for our chosen example
\begin{equation}
	\frac{\gamma_3|a_m|^2}{\gamma_1} \sim 0.1.
	\label{eq:gamma3/gamma1_value}
\end{equation}

	\subsection{Validity of the multiple scales approximation}
	\label{sec:model_validity}
In order to verify the correctness of our approximated solution achieved by multiple scales method, we compare the results of direct integration of the full motion equation \eqref{eq:motion_with_epsilon} with the steady state solution of the evolution equation \eqref{eq:steady_ampl}. We use the results from Tables~\ref{tab:eval_freq_meas_parameters} and~\ref{tab:eval_special_parameters} for $\omega_0, \alpha_3, \gamma_1$, and  $\gamma_3$. We also estimate the effective mass $m$ to be $0.7\times 10^{-12}\kilogram$, the effective capacitance to be of order of $C_0\approx 1.5\times 10^{-15}\farad$, the DC voltage $V_\text{DC}=15\volt$, the AC voltage $v=200\millivolt$, and take the distance $d$ to be the actual distance between the electrode and the mechanical beam, i.e., $d=5\micrometer$. The resulting excitation force amplitude is $f_0=600\newton\meter^{-1}$, the constant force is $F=45000\newton\meter^{-1}$ (see Eq.~\eqref{eq:motion_with_epsilon}), and the constant resonance frequency shift is $\Delta\omega_0=-2\pi\times 257\hertz$ (see Eq.~\eqref{eq:Delta_omega_0}).

\begin{figure}[htb]
    
        \includegraphics[width=3.4in]{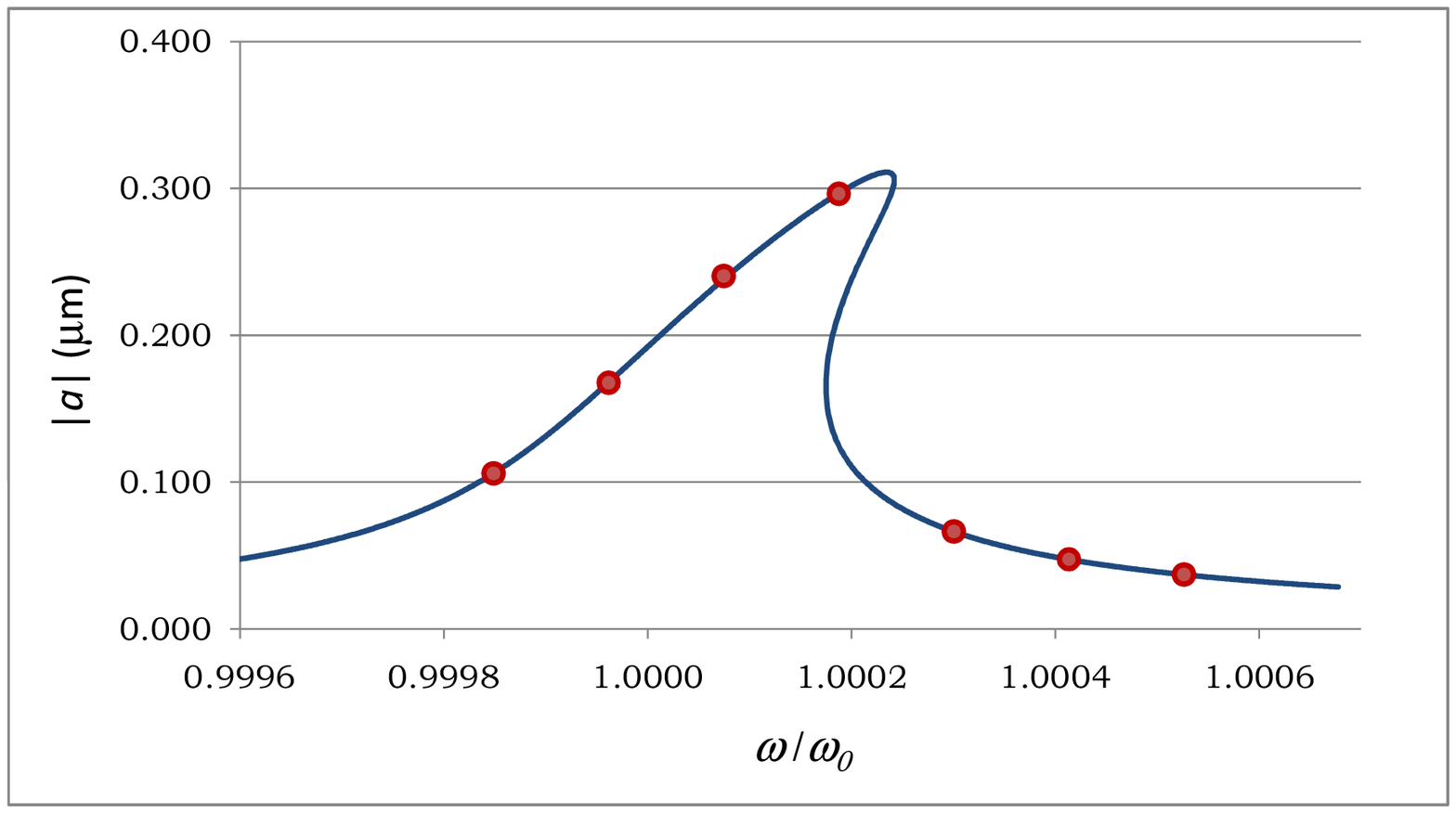}
        \caption{(Color online) Comparison of numerically calculated steady state response amplitude of the full equation of motion \eqref{eq:motion_with_epsilon} (red circles) with the steady state solution of the evolution equation \eqref{eq:steady_ampl} (solid line).}
        \label{fig:model_validity}
    
\end{figure}

In Fig.~\ref{fig:model_validity}, the exact numerical integration of Eq.~\eqref{eq:motion_with_epsilon} is compared with the solution of the approximated frequency response equation \eqref{eq:steady_ampl}. A very good correspondence between the two solutions is achieved, which validates the approximations applied in Sec.~\ref{sec:slow_env_approx}.

\section{Discussion}
\label{sec:discussion}

	\subsection{Analysis of results}
	\label{sec:results_analysis}

It follows from our experimental results that the nonlinear damping constant $p$ can be estimated with a high degree of confidence by measuring the micromechanical oscillator bistable response in the frequency domain. The values of $p$ that we find, $0.1<p<0.3$, obviously are not negligible. Referring to Eqs.~\eqref{eq:a_c} and \eqref{eq:jump_point}, we see that the considered micromechanical oscillators exhibit a damping nonlinearity that has a measurable impact on both the amplitude and frequency offset of the critical point, as well as on jump points in the bistable region. On the other hand, these values are significantly smaller then the critical value $p=1/\sqrt{3}\approx 0.577$, which would prevent the system from exhibiting bistable behavior.

Two methods of estimating the value of $p$ from frequency domain measurements were used. The first is based on a single measurement of the critical point and provides a simple means for estimating the value of $p$ by experimentally measuring the linear quality factor $Q$ at low excitation amplitude and the critical frequency shift $\Delta\omega_c$ only (see Eq.~\eqref{eq:p_from_fc}). The second can be used for any excitation amplitude that drives the system into bistable regime, but requires a comparison of different response amplitudes (see Eq.~\eqref{eq:p_quadratic_eq}). Both these methods yield similar results, however, the second one, although being less accurate, allows the experimentalist to estimate when the limit of hard excitation \cite{Nayfeh_Mook_book_95} is approached and the first order multiple scales analysis used in this study becomes inadequate. In this limit of strong excitation, the extracted values of $p$ start to diverge significantly from the results obtained at low excitation amplitudes. Our results, especially Fig.~\ref{fig:experiment_p_extract}, and the analysis of the validity of our approximations, which was carried out in Sec.~\ref{sec:model_validity}, suggest that the analysis method employed by us is adequate for a wide range of excitation amplitudes.

The third method described above allows one to estimate the value of $p$ from time domain measurements of the free ring down of the micromechanical beam oscillator based on Eq.~\eqref{eq:A_m^2_ringdown}. Although fitting results of time domain measurements to a theoretical curve introduces large inaccuracy, this method is invaluable in cases where the bistable regime cannot be achieved, e.g., due to prohibitively large amplitudes involved and the risk of pull-in.

By using the approximations developed in Sec.~\ref{sec:parameter_extraction}, we were able to estimate different parameters describing the slow envelope dynamics of our oscillators, summarized in Table~\ref{tab:eval_freq_meas_parameters}. The most important and, as far as we know, novel result is the direct estimation of the slowing down time $\tau_\text{sd}$ that is given by Eq.~\eqref{eq:tau_sd_approx}, which governs the system's dynamics in the vicinity of bifurcation point. In turn, this result is used to quantitatively evaluate the error introduced to the frequency measurements by the slowing down process, $\delta_\text{err}$ that is given by Eq.~\eqref{eq:delta_err}, which in the example studied is $2\hertz$. It can be seen that even slow sweeping rate (as compared to quasistatic rate in the linear case, which is of order of one resonant width per ring down time) can introduce a significant inaccuracy in the measured response of a micromechanical beam oscillator near bifurcation points. In our case, the inaccuracy in $\Delta\omega_j$ is about 3\%, but the inaccuracy in $\Delta\omega_j-\Delta\omega_m$ is probably much larger.

The nonlinear damping constant $p$ plays an important role in all the dynamical parameters. In the value of $K$ that isgiven by Eq.~\eqref{eq:approx_K} in our example, $p$-dependent term constitutes about 30\% of the value. The same is true for other parameters as well.

Also, we make order of magnitude estimations of thermal escape time $\tau_\text{thermal}$ (see Eq.~\eqref{eq:tau_therm_approx}), $\alpha_3$, and $\gamma_3$ (see Eq.~\eqref{eq:alpha3_gamma3_approx}), which are summarized in Table~\ref{tab:eval_special_parameters}. These approximations can be used in order to construct an accurate model of the effective one-dimensional movement of the system in the vicinity of a bifurcation point, especially if accurate enough estimations of the oscillator's amplitude and effective mass can be made.

In our case, only the order of magnitude of the parameters can be estimated. However, we were able to estimate the thermal escape time, and found the thermal escape process to be a non negligible source of inaccuracy in the frequency measurements only at very high effective noise temperatures of order $10^8\kelvin$. This result can be compared to a result from our previous work \cite{Almog_et_al_07}. In that work, a micromechanical beam oscillator similar to the ones used here was excited at a frequency between the bifurcation points. The intensity of voltage noise needed to cause transitions between these stable states was found to be $\approx 500\millivolt$, with noise bandwidth of 10\megahertz. The resulting voltage noise density is $158\microvolt/\sqrt{\hertz}$, which corresponds to an effective noise temperature $\sim 10^{13}\kelvin$. In the case of thermal escape described here, the two stable states are highly asymmetrical. The effective noise temperature of $10^8\kelvin$, which invalidates the estimations of very slow thermal escape rate in Sec.~\ref{sec:results_parameter_extraction}, corresponds to voltage noise density of $0.5\microvolt/\sqrt{\hertz}$, giving the total voltage noise intensity of 1.6\millivolt.

Finally, we can also estimate the relative contribution of the nonlinear damping term $\gamma_3|a|^2$ in the evolution equation~\eqref{eq:evolution}, and find it to be a non negligible one-tenth of the linear term $\gamma_1$ (see Eq.~\eqref{eq:gamma3/gamma1_value}) at the estimated amplitude of $|a|=100\nanometer$.

	\subsection{Geometric nonlinearities as a source of nonlinear damping}
	\label{sec:nonlin_damping_source}
	
The nature of nonlinear damping is not discussed in this work. However, nonlinear damping can be, in part, closely related to material behavior with a linear dissipation law that operates within a geometrically nonlinear regime. Here, we investigate one possible mechanism, originating from a Voigt-Kelvin type of dissipation model which describes internal viscoelastic damping in the form of a parallel spring and dashpot.

Before we proceed to build the model, one technical remark is in order. The notations in this section follow the standard ones used in continuum mechanics, and some parameters used above are redefined below. However, the end results are brought back to the form of \eqref{eq:full_eq_of_motion}.

Following Leamy and Gottlieb \cite{Leamy&Gottlieb_00, Leamy&Gottlieb_01}, we consider a planar weakly nonlinear pretensioned, viscoelastic string augmented by linear Euler-Bernoulli bending, which incorporates a Voigt-Kelvin constitutive relationship where the stress is a linear function of the strain and strain rate \cite{Meirovitch_book_97, Zener_book_48}:
\begin{equation*}
	\sigma=E\varepsilon+D\varepsilon_t,
\end{equation*}
where $\sigma$ is the stress, $\varepsilon$ is the strain, $E$ is the material Young modulus, $D$ is a viscoelastic damping parameter, and subscripts denote differentiation with respect to the corresponding variable. The equations of motion of the beam-string are
\begin{subequations}
\begin{multline}
	\rho A\tilde u_{tt}-\biggl[N\tilde u_{\tilde s}+EA\left(\tilde u_{\tilde s}+\frac12\tilde w^2_{\tilde s}\right)\\
	+DA\left(\tilde u_{t\tilde s}+\tilde w_{\tilde s}\tilde w_{t\tilde s}\right)\biggl]_{\tilde s}=0,
\end{multline}
\begin{multline}
	\rho A\tilde w_{tt}-\biggl[N\tilde w_{\tilde s}+EA\tilde w_{\tilde s}\left(\tilde u_{\tilde s}+\frac12\tilde w^2_{\tilde s}\right)\\
	+DA\tilde w_{\tilde s}\left(\tilde u_{t\tilde s}+\tilde w_{\tilde s}\tilde w_{t\tilde s}\right)-\left(EI\tilde w_{\tilde s\tilde s\tilde s}+DI\tilde w_{t\tilde s\tilde s\tilde s}\right)\biggl]_{\tilde s}=Q_{\tilde w},
\end{multline}
	\label{eq:string_eq_of_motion_orig}
\end{subequations}
where $N$ is the pretension, $\rho$ is the material density, $\tilde s $ is the material coordinate along the beam, $A$ and $I$ are the elastic element cross-sectional area and moment of inertia, respectively. Also, $\tilde u(\tilde s,t)$ and $\tilde w(\tilde s,t)$ are the respective longitudinal and transverse components of an elastic field. The generalized transverse force component $Q_{\tilde w}$ is due to external electrodynamic actuation. Note that for a parallel plate approximation,
\begin{equation*}
	Q_{\tilde w}=B\frac{[V_\text{DC}+v\cos(\omega t)]^2}{(d-\tilde w)^2},
\end{equation*}
where $V_\text{DC}$, $v$, $d$ and $\omega$ are as those defined in Eq.~\eqref{eq:full_eq_of_motion}, and $B$ is a proportionality coefficient dependent on the exact geometry of the mechanical oscillator.

We rescale the elastic field components $\tilde u$ and $\tilde w$, and the material coordinate $\tilde s$ by the beam length $L$, and time by the pretension $\sqrt{\rho AL^2/N}$ to yield a coupled set of dimensionless partial differential equations for the beam-string:
\begin{subequations}
\begin{equation}
	u_{\tau\tau}-\left[u_s+\beta\left(u_s+\frac12 w^2_s\right)+\delta\left(u_{\tau s}+w_sw_{\tau s}\right)\right]_s=0, \label {eq:string_eq_of_motion_nondim_longitudinal}
\end{equation}
\begin{multline}
	w_{\tau\tau}-\biggl[w_s+\beta w_s\left(u_s+\frac12 w^2_s\right)+\delta w_s\left(u_{\tau s}+w_sw_{\tau s}\right) \\
	-\left(\alpha w_{sss}+\mu\delta w_{\tau sss}\right)\biggl]_s=Q_w,
	\label{eq:string_eq_of_motion_nondim}
\end{multline}
\end{subequations}
where $u=\tilde u/L$, $w=\tilde w/L$, $s=\tilde s/L$ and
\begin{align*}
	\tau & =\omega_s t, \\
	\omega_s^2 & =\frac{N}{\rho AL^2}.
\end{align*}
Other dimensionless parameters include the effects of weak bending $\alpha<1$, a strong nonlinear pretension $\beta>1$, a small slenderness ratio $\mu<1$ (because $r/L\ll 1$, where $r=\sqrt{I/A}$ is the beam-string radius of gyration \cite{Meirovitch_book_97}), and finite viscoelastic damping $\delta$:
\begin{equation}
	\alpha=\frac{EI}{NL^2},\: \beta=\frac{EA}{N},\: \mu=\frac{I}{AL^2},\: \delta=\frac{D}{L}\sqrt{\frac{\beta}{\rho E}}.
	\label{eq:dimensionless_params_defs}
\end{equation}
Note that $\sqrt{\beta}$ defines the ratio between the longitudinal and transverse wave speeds \cite{Nayfeh_Mook_book_95, Meirovitch_book_97}. The rescaled parallel plate approximation is thus:
\begin{equation*}
	Q_w=\eta\frac{\left[1+\epsilon\cos(\Omega t)\right]^2}{(\gamma-w)^2},
\end{equation*}
where
\begin{equation*}
	\eta=\frac{BV_\text{DC}^2}{LN},\: \Omega=\frac{\omega}{\omega_s},\: \epsilon=\frac{v}{V_\text{DC}},\: \gamma=\frac{d}{L}.
\end{equation*}

We note that as the first longitudinal natural frequency is much higher than the first transverse natural frequency ($\beta\gg1$), the longitudinal inertia and damping terms in Eq.~\eqref{eq:string_eq_of_motion_nondim_longitudinal} can be neglected to yield a simple spatial relationship between the transverse and longitudinal derivatives. Incorporating fixed boundary conditions ($u(0,\tau)=u(1,\tau)=0)$ enables integration of the resulting relationship to yield:
\begin{equation*}
	u_s=-\frac12 w^2_s+c_1(\tau),
\end{equation*}
where
\begin{equation*}
	c_1=\frac12\int_0^1 w^2_s ds.
\end{equation*}
Thus, the resulting weakly nonlinear beam-string initial boundary value problem consists of an integro-differential equation for the transverse mode:
\begin{multline}
	w_{\tau\tau}-w_{ss}\left(1+\beta c_1(\tau)+\delta c_{1\tau}(\tau)\right) \\
	+\alpha w_{ssss}+\mu\delta w_{\tau ssss}=Q_w,
	\label{eq:integro-differential_beam}
\end{multline}
where
\begin{equation*}
	c_{1\tau}=\int^1_0w_sw_{\tau s}ds.
\end{equation*}
In order to facilitate comparison of the continuum model with the lumped mass model in Eq.~\eqref{eq:full_eq_of_motion}, we consider a localized electrodynamic force $Q_w=Q_w(s=1/2,\tau)$.

We reduce the integro-differential field equation in \eqref{eq:integro-differential_beam} and its fixed boundary conditions to a modal dynamical system via an assumed single mode Galerkin assumption, $w(s,\tau)=q_1(\tau)\phi_1(s)$, using a harmonic string mode $\phi_1=\sqrt{2}\sin(\pi s)$:
\begin{multline}
	I_1 q_{\tau\tau}-I_2 q\left[1+I_3\left(\frac12\beta q^2+\delta qq_\tau\right)\right]+I_4 \left(\alpha q+\mu\delta q_\tau\right)\\
	=I_5 \eta\frac{\left[1+\epsilon\cos(\Omega\tau)\right]^2}{\left(\gamma-I_5 q\right)^2},
\end{multline}
where $q=q_1$ and the integral coefficients are:
\begin{align*}
	&I_1=\int_0^1\phi_1^2ds=1,\\
	&I_2=\int_0^1\phi_1\phi_{1ss}ds=-\pi^2,\\
	&I_3=\int_0^1\phi_{1s}^2ds=\pi^2,\\
	&I_4=\int_0^1\phi_1\phi_{1ssss}ds=\pi^4,\\
	&I_5=\phi_1\left(\frac12\right)=\sqrt{2}.
\end{align*}

It is convenient to rescale the maximal response $|w(1/2,\tau)|=q\bar\phi$, where $\bar\phi=\phi_1(1/2)=\sqrt{2}$, by the dimensionless gap $z=q\bar\phi/\gamma$, and to rescale time by the unperturbed $(\eta=0)$ natural frequency $t'=\tilde\omega_1\tau$, where $\tilde\omega_1=\sqrt{\alpha I_4-I_2}=\pi\sqrt{1+\alpha\pi^2}$. The resulting dynamical system is:
\begin{equation}
	\ddot z+\left(\frac{1}{Q}+\hat\delta z^2\right)\dot z + \left(1+\hat\beta z^2\right)z=\hat\eta\frac{[1+\epsilon\cos(\hat\Omega t')]^2}{(1-z)^2},
	\label{eq:string_eq_of_motion_one_dim}
\end{equation}
where
\begin{align*}
	\hat\beta & =\frac{|I_2|I_3\beta\gamma^2}{2\bar\phi^2\tilde\omega_1^2},\\
	\frac{1}{Q} & =\frac{I_4\mu\delta}{\tilde\omega_1},\\
	\hat\delta & =\frac{\delta|I_2|I_3\gamma^2}{\bar\phi^2\tilde\omega_1},\\
	\hat\eta & =\frac{I_5\eta\bar\phi}{\gamma^3\tilde\omega_1^2},\\
	\hat\Omega & =\frac{\Omega}{\tilde\omega_1}.
\end{align*}
Note that the ratio between nonlinear and linear damping in Eq. \eqref{eq:string_eq_of_motion_one_dim} consists of only the beam-string geometric properties \cite{Mintz_thesis_09}. For example, a typical ratio is $\hat\delta Q=6d^2/h^2\approx 65$ for a beam-string with a prismatic cross-section, where $h=1.5\micrometer$ is the dimension of the beam-string in the transverse direction $w$, and $d=5\micrometer$ is the resonator gap.

The last equation \eqref{eq:string_eq_of_motion_one_dim} can be compared, after rescaling, to the dimensional equation \eqref{eq:full_eq_of_motion}, which we rewrite here for convenience after some rearrangement and simplification (e.g. $\gamma_{32}=0$):
\begin{multline}
    \ddot x+(2\gamma_{11}+\gamma_{31}x^2)\dot x+(\omega_0^2+\alpha_3x^2)x\\
	=\frac{C_0V_\text{DC}^2}{2md}\frac{\left[1+\frac{v}{V_\text{DC}}\cos\omega t\right]^2}{\left(1-\frac{x}{d} \right)^2}.
     \label{eq:full_eq_of_motion_for_comparison}
\end{multline}
The comparison of Eq.~\eqref{eq:string_eq_of_motion_one_dim} with Eq.~\eqref{eq:full_eq_of_motion_for_comparison} results in:
\begin{subequations}
	\begin{align}
		&  \alpha_{3}=\frac{\hat\beta\omega_0^2}{d^2}=\frac{\pi^2 EA\omega_0^2}{4NL^2}=\frac{\pi^2\alpha A\omega_0^2}{4I},\\
		&  \gamma_{11}=\frac{\omega_0}{2Q},\\
		&  \gamma_{31}=\frac{\hat\delta\omega_0}{d^2}=\frac{A\gamma_{11}}{I},\\
		&  F=\frac{C_0 V_\text{DC}^2}{2md}=\hat\eta d\omega_0^2.
	\end{align}
	\label{eq:param_estimations_continuous_model}%
\end{subequations}
The last results can be used to estimate the lower bound of nonlinear damping due to nonlinear pretension of a viscoelastic string. Using Eqs.~\eqref{eq:p}, \eqref{eq:dimensionless_params_defs}, \eqref{eq:param_estimations_continuous_model}, and
\begin{equation*}
	I  =\frac{Ah^{2}}{12}
\end{equation*}
for prismatic cross-section, one has
\begin{equation}
	p_{\text{min}}=\frac{2}{3}\frac{\tilde{\omega}_{1}\delta}{\beta}\approx\frac{8}{\pi^{2}}\frac{1}{Q}\left( \frac{L}{h}\right)  ^{2}\frac{N}{EA}=\frac{2}{3\pi^2 Q\alpha},
	\label{eq:p_estimation_cont_model}
\end{equation}
where $h$ denotes the dimension of the beam-string in the transverse direction $w$.

It is possible to estimate the order of magnitude of $p_\text{min}$ in Eq.~\eqref{eq:p_estimation_cont_model} for metals using the fact that the Young modulus of bulk metals $E\sim O(10^{10})\div O(10^{11})$ \pascal. Also, the largest value of $N/A$ that is still compatible with elastic behavior can be approximated by half the ultimate tensile strength, which is about $50\div 100\times10^6$ \pascal for most metals. For our beam-strings discussed above, $L= 100\div200\micrometer, h\approx 1\micrometer$. Using these values results in $p\sim O(10^{-4})\div O(10^{-3})$. For longer and wider beams ($L=500\micrometer, h=1.5\micrometer$) fabricated and measured using the same methods \cite{Mintz_thesis_09}, the lower bound on nonlinear damping coefficient given by Eq.~\eqref{eq:p_estimation_cont_model} is $p_\text{min}\sim 0.0022\div 0.045$, while the range of values extracted from the experiment is $0.015<p<0.151$ \cite{Mintz_thesis_09}. Although the elastic properties of a specific metal or alloy used in micro machined devices might differ significantly from the bulk values, they are still likely to fall inside the ranges defined above. Therefore, a linear viscoelastic process with a pure Voigt-Kelvin dissipation model can serve as a possible lower bound but cannot account for the main part of nonlinear dissipation rate found in our experiments.

Unfortunately, theory describing the processes underlying nonlinear damping in micromechanical beam is virtually non-existent at this moment, and no clear tendencies in the value of $p$ were observed during the experiments. Therefore, the exact behavior of nonlinear damping term during beam scaling and its dependence on the linear $Q$ of the structure remains elusive. Further experiments with wider range of micromechanical beams are needed to establish this behavior and to pinpoint the most significant mechanisms of dissipation.

\section{Summary}

In this study, the nonlinear dynamical behavior of an electrically excited micromechanical doubly clamped beam oscillator was investigated in vacuum. The micromechanical beam was modeled as a Duffing-like single degree of freedom oscillator, nonlinearly coupled to a thermal bath. Using the method of multiple scales, we were able to construct a detailed model of slow envelope behavior of the system, including effective noise terms.

It follows from the model that nonlinear damping plays an important role in the dynamics of the micromechanical beam oscillator. Several methods for experimental evaluation of the nonlinear damping contribution were proposed, applicable at different experimental situations. These methods were compared experimentally and shown to provide similar results. The experimental values of the nonlinear damping constant are non negligible for all the beams measured.

Also, the slow envelope model was used to describe the behavior of the system close to bifurcation points in the presence of nonlinear damping. In the vicinity of these points, the dynamics of the system is significantly slowed down, and the phase plane motion becomes essentially one-dimensional. We have defined several parameters that govern the dynamics of the micromechanical beam oscillator in these conditions, and have provided simple approximations that can be used to estimate these parameters from experimental data.

The approximations developed in this study can be utilized by the experimentalist in order to estimate the inaccuracy of frequency response measurements of Duffing-like oscillators in the vicinity of bifurcation points. Applying these results to our samples, we have found that thermal escape process near the bifurcation point causes measurement inaccuracy that is negligible. In contrast, the slowing down phenomenon, which is a characteristic of saddle-node bifurcation, can contribute a significant error to the measured frequency response. This error is non negligible even at relatively slow frequency sweeping rates. Similar methods can be utilized for other parameter sweeping measurements, such as excitation amplitude sweeping.

As part of an effort to explain the origins of the nonlinear damping, we have formulated and analyzed a model of a planar, weakly nonlinear pretensioned, viscoelastic string augmented by linear Euler-Bernoulli bending, which incorporates a Voigt-Kelvin constitutive relationship. This model exemplifies one of the possible causes of non negligible nonlinear damping observed in the experiment. Based on this model, we have determined a simple relation connecting the maximal expected value of the nonlinear damping parameter, the bulk Young modulus of the material, and its yielding stress. However, while this model can serve as a lower bound, it cannot account for the full magnitude of the nonlinear damping measured in the experiment. Additional experimental and theoretical work is required to enhance our understanding of the phenomenon of nonlinear damping in microelectromechanical systems.

In this work we have demonstrated conclusively that nonlinear damping in micromechanical doubly-clamped beam oscillator may play an important role. The methods presented in this paper may allow a systematic study of nonlinear damping in micro- and nanomechanical oscillators, which may help revealing the underlying physical mechanisms.

\begin{acknowledgments}
We would like to thank R. Lifshitz for many fruitful discussions. This work was partially supported by Intel Corporation, the Israeli Ministry of Science, the Israel Science foundation, the German Israel foundation, and the Russell Berry foundation.
\end{acknowledgments}

\bibliographystyle{unsrt}

\end{document}